%% file: paper.tex
\documentclass[fleqn,usenatbib]{mnras}

\usepackage{newtxtext,newtxmath}

\usepackage[T1]{fontenc}

\DeclareRobustCommand{\VAN}[3]{#2}
\let\VANthebibliography\thebibliography
\def\thebibliography{\DeclareRobustCommand{\VAN}[3]{##3}\VANthebibliography}

\usepackage{graphicx}	
\usepackage{amsmath}	

\usepackage{array}

\input{affils}

\newcommand{\M}{M$_\odot$}
\newcommand{\kms}{km\,s$^{-1}$}

\title[Physical properties of TDE sub-classes]{Systematic light curve modelling of TDEs: statistical differences between the spectroscopic classes}

\author[M.~Nicholl et al]{{Matt Nicholl$^{1,2}$}\thanks{Contact e-mail: \href{mailto:m.nicholl.1@bham.ac.uk}{m.nicholl.1@bham.ac.uk}}, Daniel Lanning$^1$, Paige Ramsden$^1$, Brenna Mockler$^{3,4}$, Andy Lawrence$^{5}$, Phil Short$^{5}$,  \newauthor Evan J.~Ridley$^{1,2}$
\\
$^{1}$\bham\\
$^{2}$\igw\\
$^{3}$\ucsc\\
$^{4}$\dark\\
$^{5}$\edinburgh
}

\date{Accepted XXX. Received YYY; in original form ZZZ}

\pubyear{2021}

\begin{document}
\label{firstpage}
\pagerange{\pageref{firstpage}--\pageref{lastpage}}
\maketitle

\begin{abstract}


    With the sample of observed tidal disruption events (TDEs) now reaching several tens, distinct spectroscopic classes have emerged: TDEs with only hydrogen lines (TDE-H), only helium lines (TDE-He), or hydrogen in combination with He\,II and often N\,III/O\,III (TDE-H+He). Here we model the light curves of 32 optically-bright TDEs using the Modular Open Source Fitter for Transients (\textsc{mosfit}) to estimate physical and orbital properties, and look for statistical differences between the spectroscopic classes. For all types, we find a shallow distribution of star masses, compared to a typical initial mass function, between $\sim 0.1-1$\,\M, and no TDEs with very deep encounters. Our main result is that TDE-H events appear to come from less complete disruptions (and possibly lower SMBH masses) than TDE-H+He, with TDE-He events fully disrupted. We also find that TDE-H events have more extended photospheres, in agreement with recent literature, and argue that this could be a consequence of differences in the self-intersection radii of the debris streams. Finally, we identify an approximately linear correlation between black hole mass and radiative efficiency. We suggest that TDE-H may be powered by collision-induced outflows at relatively large radii, while TDE-H+He could result from prompt accretion disks, formed more efficiently in closer encounters around more massive SMBHs.
     
     \end{abstract}




\section{Introduction}

A tidal disruption event (TDE) is a luminous flare resulting from the destruction of a star by a supermassive black hole (SMBH). It occurs when the orbit of the star intersects the tidal radius, where the strong gradient in the gravitational force from the SMBH overwhelms the self-gravity of the star, producing streams of bound and unbound debris. For a canonical `full' disruption, the bound stream comprises about half the mass of the original star  \citep[e.g.][]{Hills1975,Rees1988,Phinney1989,Kochanek1994,Guillochon2013}. Relativistic precession of the orbit allows the stream to self-intersect and liberate energy, either directly in these stream collisions \citep{Piran2015,Jiang2016} or through subsequent accretion onto the SMBH.

Emission from TDEs has been detected from X-rays \citep{Komossa2002,Auchettl2017} to radio \citep{Alexander2020}, and even potentially in neutrinos \citep{Stein2021}, but most TDEs are discovered first at optical wavelengths by wide-field photometric transient surveys. TDEs detected in this way have a colour temperature of a few $\times 10^4$\,K, an order of magnitude lower than expected for a compact accretion disk. This suggests that the light originates -- or is reprocessed in -- an extended photosphere or debris atmosphere \citep{Loeb1997,Strubbe2009,Roth2016,Dai2018}.

\begin{figure}
    \centering
    \includegraphics[width=\columnwidth]{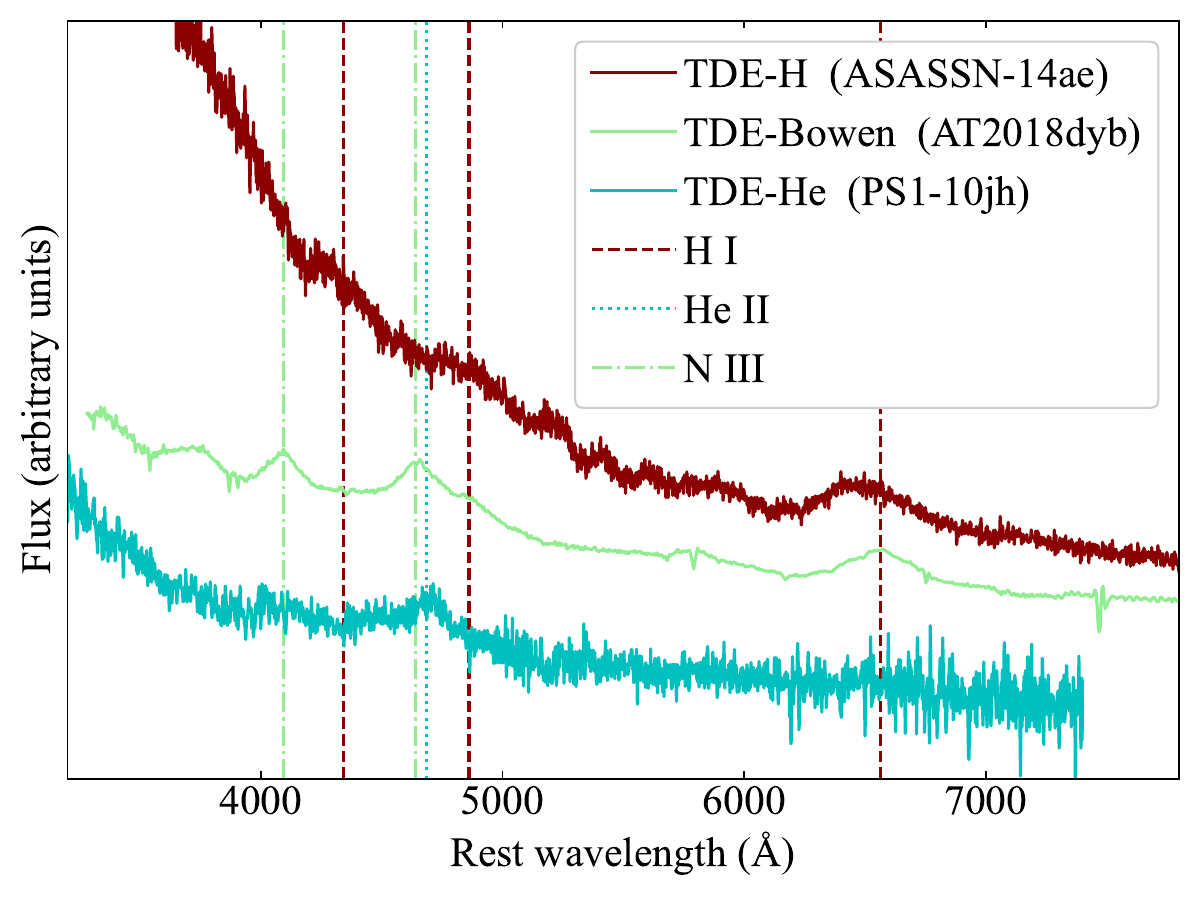}
    \caption{Examples of the main TDE spectroscopic classes. Defining emission lines are marked with vertical lines. Data are from \citet{Gezari2012}, \citet{Holoien2014} and \citet{Leloudas2019}}
    \label{fig:spec}
\end{figure}

Spectroscopic studies have revealed significant diversity within this optically-discovered population. Most TDEs exhibit broad emission lines with widths of a few $100-1000$\,\kms, the strongest of which were quickly identified as H\,I and/or He\,II \citep{Gezari2012,Arcavi2014,Holoien2016a}. However, recent studies have revealed TDEs with emission from N\,III, consistent with the Bowen Fluorescence mechanism \citep{Blagorodnova2019,Leloudas2019,Onori2019}, and Fe\,II lines indicative of a low-ionisation disk \citep{Wevers2019}. The line profiles are also very heterogeneous, with some showing clear disk-like double-peaks \citep{Short2020,Hung2020}. Spectroscopic signatures of outflows have been detected for some TDEs in the UV \citep{Hung2019,Hung2021} and optical \citep{Roth2018,Nicholl2020}, in some cases only appearing months after the peak of the flare \citep{Nicholl2019}. But perhaps the most important diversity is in the line ratios, with about half of TDEs showing almost exclusively H\,I lines, a small number only He\,II, and the rest a mixture of H\,I, He\,II and N\,III \citep{Arcavi2014,Leloudas2019,vanVelzen2021a}. In the parlance of our times, these are now the spectroscopic sub-classes TDE-H, TDE-He, and TDE-H+He (originally TDE-Bowen) \citep{vanVelzen2021a}. Examples of each class are shown in Figure \ref{fig:spec}.

Given the bolometric luminosity or a model for the reprocessed spectral energy distribution \citep[e.g.][]{Loeb1997,Metzger2016,Lu2020}, the light curves of TDEs can be decoded to reveal details of the disruption. The Eddington limit, a function of SMBH mass, provides a natural luminosity scale, while the peak mass return rate and duration scale with the mass of the star \citep[e.g.][]{Ulmer1999,Guillochon2013}. The time evolution of the fallback rate depends on how complete was the disruption, and therefore how deeply penetrating was the orbit \citep{Lodato2009,Guillochon2013,Coughlin2019}. \citet{Mockler2019}, following \citet{Guillochon2014}, presented an analytic model that captures these essential scaling relations, calibrated to numerical simulations \citep{Guillochon2013}, within the Modular Open Source Fitter for Transients \citep[\textsc{mosfit};][]{Guillochon2018}, and used this to fit for the system parameters of 14 TDEs -- the bulk of the optically-detected population at that time.

As we enter the statistical age of TDE science, with tens of TDEs now discovered by surveys like ZTF, PTF, PanSTARRS, ASAS-SN and ATLAS -- and the imminent Vera Rubin Observatory predicted to find thousands of TDEs \citep{Bricman2020} -- we can look for differences in their derived physical properties that correlate with spectroscopic typing. \citet{vanVelzen2021a} showed from their statistical sample that the TDE-H+He objects typically have smaller photospheric radii than objects with a TDE-H spectrum, and connected this finding to the high densities required for efficient operation of the Bowen fluorescence mechanism. \citet{Hinkle2020} found a similar difference in radii between these classes, while finding no differences in the light curve slopes. \citet{Charalampopoulos2021} found a clear correlation between H$\alpha$ luminosity and blackbody radius, as well as finding lower line velocities in the TDE-H+He sample. \citet{Mockler2021b} recently showed that TDE-H+He events may also result preferentially from stars more massive than 1.3\,\M.

In this paper, we study 32 TDEs with well-observed light curves, and fit these with the \textsc{mosfit} TDE model. Examining the derived physical parameters for each spectroscopic sub-class, we find statistically significant differences not only in photospheric radii, but also in the impact parameter, with TDE-H+He and TDE-He events resulting from more complete disruptions, and tentative evidence that these occur around more massive SMBHs. We also present the mass distribution of disrupted stars, and new evidence for a correlation between SMBH mass and radiative efficiency. Host galaxy bulge masses for most of the TDE sample analysed here, and their correlation with SMBH mass, are presented in a companion paper \citep{Ramsden2022}. We describe our model and data selection in section \ref{sec:analysis}, with the statistical results in section \ref{sec:results}. We discuss the physical implications in section \ref{sec:disc} and conclude in section \ref{sec:conc}.

\section{Data analysis}
\label{sec:analysis}

\subsection{Mosfit model}
\label{sec:model}

The \textsc{mosfit} TDE model has nine free parameters: the masses of the SMBH ($M_\bullet$) and disrupted star ($M_*$); a scaled impact parameter ($b$) which determines whether the star is fully or partially disrupted; the efficiency of converting mass fallback to radiation ($\epsilon$); the normalisation of the photospheric radius ($R_{\rm ph,0}$) and a power-law index ($l_{\rm ph}$) relating this radius to the instantaneous luminosity; the viscous delay time ($t_\nu$) between fallback and production of radiation (e.g. due to suspension in a disk); the time {relative to the first detection at which the light curve begins} ($t_0$); the extinction in the host galaxy, proportional to the column density ($n_{\rm H,host}$); and a white noise parameter ($\sigma$).

\begin{table*}
\caption{Full TDE sample, with medians and 16th/84th percentiles of the marginalised posteriors from the \textsc{mosfit} model parameters. The parameters are described in section \ref{sec:model}, with the priors listed in Table \ref{tab:priors}. {We do not list the $t_0$ (start time) or $\sigma$ (white noise) parameters as they have no physical significance.}}
  \footnotesize
  \centering
\begin{tabular}{ccccccccccc}
TDE & Type$^a$ & $\log(M_{\bullet}/{\rm M}_{\odot})$ & $M_{*}/{\rm M}_{\odot}$ & $b$ & $\log(\epsilon)$ & $\log(R_{\rm ph,0})$ & $l_{\rm ph}$ & $\log(T_{\nu})$& $\log(n_{\rm H})$  & Data \\
\hline 
ASASSN-14ae & H          & 6.13$^{+0.05}_{-0.04}$ & 0.99$^{+0.14}_{-0.08}$ & 0.27$^{+0.04}_{-0.05}$ & -1.9$^{+0.22}_{-0.24}$ & 2.41$^{+0.14}_{-0.15}$ & 1.54$^{+0.12}_{-0.12}$ & -0.6$^{+0.75}_{-1.13}$ & 18.7$^{+1.05}_{-1.66}$  & 1,2 \\
ASASSN-14li$^b$ & H+He$^*$      & 7.00$^{+0.08}_{-0.11}$ & 0.18$^{+0.07}_{-0.05}$ & 0.87$^{+0.09}_{-0.09}$ & -1.0$^{+0.17}_{-0.17}$ & 0.46$^{+0.22}_{-0.19}$ & 1.88$^{+0.14}_{-0.18}$ & -1.3$^{+1.30}_{-1.04}$ & 20.7$^{+0.03}_{-0.04}$  & 3,2 \\
ASASSN-15oi & He$^*$         & 6.73$^{+0.02}_{-0.02}$ & 0.12$^{+0.11}_{-0.02}$ & 1.03$^{+0.02}_{-0.02}$ & -2.1$^{+0.09}_{-0.18}$ & 2.73$^{+0.24}_{-0.25}$ & 2.55$^{+0.17}_{-0.15}$ & -2.0$^{+0.93}_{-0.58}$ & 19.5$^{+0.45}_{-0.86}$  & 4,2 \\
AT2017eqx$^c$ & H+He     & 6.56$^{+0.08}_{-0.09}$ & 0.11$^{+0.08}_{-0.01}$ & 0.98$^{+0.08}_{-0.11}$ & -1.5$^{+0.27}_{-0.32}$ & 1.07$^{+0.22}_{-0.19}$ & 1.56$^{+0.11}_{-0.09}$ & -1.3$^{+1.08}_{-1.08}$ & 20.9$^{+0.08}_{-0.12}$  & 5,2 \\
AT2018dyb & H+He         & 7.19$^{+0.02}_{-0.03}$ & 0.10$^{+0.00}_{-0.00}$ & 0.99$^{+0.01}_{-0.02}$ & -0.6$^{+0.05}_{-0.06}$ & 0.95$^{+0.08}_{-0.07}$ & 2.12$^{+0.03}_{-0.04}$ & -1.8$^{+0.87}_{-0.45}$ & 20.8$^{+0.01}_{-0.01}$  & 6,7,2 \\
AT2018hco & H            & 6.64$^{+0.14}_{-0.15}$ & 0.08$^{+0.01}_{-0.01}$ & 0.84$^{+0.15}_{-0.10}$ & -1.1$^{+0.16}_{-0.13}$ & 0.50$^{+0.07}_{-0.11}$ & 0.79$^{+0.09}_{-0.09}$ & -0.2$^{+0.97}_{-1.96}$ & 20.0$^{+0.48}_{-1.75}$  & 8,2 \\
AT2018hyz$^d$ & H+He$^*$     & 6.57$^{+0.04}_{-0.04}$ & 0.96$^{+0.04}_{-0.05}$ & 0.88$^{+0.11}_{-0.04}$ & -2.6$^{+0.06}_{-0.08}$ & 1.47$^{+0.07}_{-0.08}$ & 1.08$^{+0.05}_{-0.06}$ & -1.3$^{+1.16}_{-1.00}$ & 19.3$^{+0.66}_{-1.48}$  & 7,8,2 \\
AT2018iih & He           & 6.92$^{+0.02}_{-0.02}$ & 0.96$^{+0.03}_{-0.04}$ & 0.95$^{+0.00}_{-0.00}$ & -1.7$^{+0.02}_{-0.01}$ & 0.46$^{+0.02}_{-0.02}$ & 0.36$^{+0.02}_{-0.02}$ & -0.7$^{+0.90}_{-1.31}$ & 17.5$^{+1.35}_{-0.98}$  & 9,2 \\
AT2018lna & H+He         & 6.67$^{+0.13}_{-0.12}$ & 0.18$^{+0.14}_{-0.07}$ & 0.87$^{+0.12}_{-0.15}$ & -0.7$^{+0.21}_{-0.29}$ & -0.0$^{+0.10}_{-0.09}$ & 1.84$^{+0.65}_{-0.44}$ & -0.8$^{+1.22}_{-1.33}$ & 17.8$^{+1.18}_{-1.14}$  & 8,2 \\
AT2018zr & H$^*$             & 6.79$^{+0.04}_{-0.04}$ & 0.22$^{+0.11}_{-0.08}$ & 0.83$^{+0.07}_{-0.08}$ & -2.2$^{+0.19}_{-0.22}$ & 3.13$^{+0.22}_{-0.23}$ & 2.05$^{+0.13}_{-0.13}$ & -1.7$^{+0.90}_{-0.78}$ & 19.7$^{+0.41}_{-1.00}$  & 8,2 \\
AT2019ahk & H$^*$            & 6.71$^{+0.08}_{-0.11}$ & 0.11$^{+0.03}_{-0.03}$ & 0.78$^{+0.28}_{-0.24}$ & -1.6$^{+0.27}_{-0.17}$ & 2.22$^{+0.27}_{-0.20}$ & 1.90$^{+0.26}_{-0.24}$ & -0.4$^{+0.81}_{-1.07}$ & 19.3$^{+0.36}_{-0.51}$  & 10,2 \\
AT2019azh & H+He$^*$         & 6.70$^{+0.06}_{-0.07}$ & 0.47$^{+0.20}_{-0.07}$ & 0.43$^{+0.09}_{-0.05}$ & -1.0$^{+0.17}_{-0.30}$ & 0.20$^{+0.04}_{-0.05}$ & 0.94$^{+0.10}_{-0.10}$ & -0.1$^{+0.50}_{-0.81}$ & 20.4$^{+0.06}_{-0.09}$  & 8,2 \\
AT2019bhf & H            & 6.57$^{+0.13}_{-0.12}$ & 0.33$^{+0.44}_{-0.22}$ & 0.80$^{+0.20}_{-0.28}$ & -2.3$^{+0.62}_{-0.59}$ & 3.30$^{+0.41}_{-0.49}$ & 2.20$^{+0.26}_{-0.28}$ & -0.3$^{+0.73}_{-1.01}$ & 18.9$^{+1.13}_{-1.50}$  & 8,2 \\
AT2019cho & H+He         & 6.71$^{+0.09}_{-0.08}$ & 0.12$^{+0.06}_{-0.02}$ & 0.89$^{+0.09}_{-0.10}$ & -1.7$^{+0.11}_{-0.16}$ & 1.83$^{+0.14}_{-0.15}$ & 1.39$^{+0.14}_{-0.13}$ & -1.2$^{+1.15}_{-1.13}$ & 18.1$^{+1.23}_{-1.30}$  & 8,2 \\
AT2019dsg$^c$ & H+He$^*$         & 6.57$^{+0.21}_{-0.16}$ & 0.91$^{+0.22}_{-0.63}$ & 0.48$^{+0.09}_{-0.07}$ & -1.8$^{+0.63}_{-0.13}$ & -0.0$^{+0.09}_{-0.06}$ & 0.49$^{+0.57}_{-0.08}$ & -0.8$^{+1.07}_{-1.30}$ & 20.1$^{+0.39}_{-0.38}$  & 8,11,2 \\
AT2019ehz & H$^*$            & 6.34$^{+0.04}_{-0.04}$ & 0.10$^{+0.00}_{-0.00}$ & 0.86$^{+0.06}_{-0.07}$ & -1.2$^{+0.10}_{-0.10}$ & 1.22$^{+0.11}_{-0.09}$ & 2.43$^{+0.17}_{-0.15}$ & 0.74$^{+0.06}_{-0.07}$ & 20.7$^{+0.05}_{-0.06}$ & 8,2 \\
AT2019eve$^c$ & H        & 5.79$^{+0.08}_{-0.07}$ & 0.12$^{+0.14}_{-0.02}$ & 0.82$^{+0.15}_{-0.16}$ & -2.9$^{+0.16}_{-0.27}$ & 3.10$^{+0.56}_{-0.72}$ & 0.79$^{+0.62}_{-0.49}$ & -1.4$^{+0.90}_{-0.93}$ & 18.3$^{+1.47}_{-1.48}$  & 8,2 \\
AT2019lwu$^c$ & H        & 6.31$^{+0.13}_{-0.11}$ & 0.59$^{+0.41}_{-0.48}$ & 0.53$^{+0.19}_{-0.17}$ & -2.5$^{+0.84}_{-0.61}$ & 3.42$^{+0.36}_{-0.49}$ & 2.31$^{+0.31}_{-0.36}$ & -1.3$^{+1.11}_{-0.99}$ & 18.5$^{+1.38}_{-1.56}$  & 8,2 \\
AT2019meg & H            & 6.52$^{+0.06}_{-0.06}$ & 0.10$^{+0.00}_{-0.00}$ & 0.77$^{+0.10}_{-0.09}$ & -0.8$^{+0.18}_{-0.20}$ & 0.94$^{+0.13}_{-0.10}$ & 2.74$^{+0.54}_{-0.45}$ & -0.7$^{+0.99}_{-1.29}$ & 20.7$^{+0.07}_{-0.11}$  & 8,2 \\
AT2019mha & H            & 6.25$^{+0.12}_{-0.15}$ & 2.62$^{+1.38}_{-0.77}$ & 0.37$^{+0.06}_{-0.07}$ & -2.7$^{+0.21}_{-0.20}$ & 3.26$^{+0.35}_{-0.38}$ & 3.44$^{+0.24}_{-0.25}$ & 0.98$^{+0.09}_{-0.17}$ & 19.4$^{+0.56}_{-0.85}$  & 8,2 \\
AT2019qiz & H+He$^*$         & 6.22$^{+0.04}_{-0.04}$ & 1.01$^{+0.03}_{-0.02}$ & 0.54$^{+0.02}_{-0.02}$ & -3.2$^{+0.09}_{-0.08}$ & 1.12$^{+0.06}_{-0.06}$ & 0.69$^{+0.02}_{-0.02}$ & 0.58$^{+0.08}_{-0.13}$ & 20.5$^{+0.11}_{-0.18}$  & 8,12,2 \\
GALEX D1-9 & Unknown           & 6.79$^{+0.25}_{-0.33}$ & 0.28$^{+0.21}_{-0.14}$ & 1.01$^{+0.14}_{-0.11}$ & -1.6$^{+0.43}_{-0.23}$ & -0.9$^{+0.22}_{-0.16}$ & 0.07$^{+0.06}_{-0.04}$ & -0.4$^{+1.23}_{-1.44}$ & 18.3$^{+1.52}_{-1.46}$  & 13 \\
GALEX D3-13$^b$ & Unknown$^*$          & 7.00$^{+0.21}_{-0.22}$ & 0.36$^{+0.79}_{-0.12}$ & 0.99$^{+0.09}_{-0.10}$ & -1.6$^{+0.37}_{-0.37}$ & -0.4$^{+0.20}_{-0.18}$ & 0.55$^{+0.17}_{-0.12}$ & -0.4$^{+0.94}_{-1.27}$ & 19.8$^{+0.57}_{-1.44}$  & 13 \\
OGLE16aaa & Unknown & 6.40$^{+0.05}_{-0.04}$ & 0.18$^{+0.11}_{-0.09}$ & 1.00$^{+0.05}_{-0.06}$ & -1.4$^{+0.31}_{-0.21}$ & 0.83$^{+0.06}_{-0.07}$ & 0.83$^{+0.12}_{-0.12}$ & -0.9$^{+1.09}_{-1.20}$ & 17.8$^{+1.04}_{-1.10}$  & 14,2 \\
PS1-10jh$^b$ & He            & 7.00$^{+0.04}_{-0.06}$ & 0.41$^{+0.09}_{-0.16}$ & 0.96$^{+0.04}_{-0.07}$ & -1.8$^{+0.28}_{-0.14}$ & 0.69$^{+0.13}_{-0.12}$ & 1.25$^{+0.10}_{-0.09}$ & 0.38$^{+0.40}_{-2.09}$ & 18.7$^{+0.99}_{-1.60}$  & 15 \\
PS1-11af & Unknown   & 6.45$^{+0.05}_{-0.04}$ & 0.96$^{+0.03}_{-0.06}$ & 0.88$^{+0.08}_{-0.05}$ & -2.7$^{+0.06}_{-0.04}$ & 1.07$^{+0.07}_{-0.06}$ & 0.97$^{+0.09}_{-0.07}$ & -0.8$^{+0.99}_{-1.44}$ & 19.0$^{+0.68}_{-0.71}$  & 16 \\
PTF09djl & H             & 6.42$^{+0.11}_{-0.11}$ & 0.22$^{+0.22}_{-0.10}$ & 0.32$^{+0.18}_{-0.08}$ & -0.9$^{+0.36}_{-0.81}$ & 2.05$^{+1.28}_{-0.87}$ & 1.59$^{+0.70}_{-0.33}$ & 0.00$^{+0.44}_{-0.69}$ & 19.9$^{+0.62}_{-1.03}$  & 17 \\
PTF09ge & He             & 6.47$^{+0.04}_{-0.04}$ & 0.08$^{+0.01}_{-0.00}$ & 1.18$^{+0.06}_{-0.05}$ & -2.1$^{+0.07}_{-0.10}$ & 3.30$^{+0.16}_{-0.18}$ & 1.95$^{+0.09}_{-0.08}$ & -1.1$^{+1.17}_{-1.18}$ & 18.5$^{+0.80}_{-0.94}$  & 17 \\
SDSS TDE1 & Unknown           & 6.84$^{+0.31}_{-0.34}$ & 0.23$^{+0.19}_{-0.10}$ & 0.95$^{+0.12}_{-0.16}$ & -1.6$^{+0.47}_{-0.39}$ & 0.07$^{+0.37}_{-0.28}$ & 0.71$^{+0.29}_{-0.19}$ & 0.24$^{+0.74}_{-1.49}$ & 19.2$^{+1.19}_{-1.29}$  & 18 \\
SDSS TDE2 & H                 & 6.66$^{+0.34}_{-0.47}$ & 0.32$^{+0.39}_{-0.15}$ & 1.05$^{+0.35}_{-0.10}$ & -0.9$^{+0.32}_{-0.32}$ & 0.61$^{+0.20}_{-0.25}$ & 1.09$^{+0.72}_{-0.27}$ & -0.6$^{+1.48}_{-1.28}$ & 19.8$^{+0.66}_{-1.79}$  & 18 \\
iPTF16axa & H+He         & 7.29$^{+0.11}_{-0.13}$ & 0.33$^{+0.17}_{-0.07}$ & 0.94$^{+0.08}_{-0.11}$ & -0.6$^{+0.14}_{-0.28}$ & 0.32$^{+0.25}_{-0.18}$ & 2.30$^{+0.32}_{-0.40}$ & 0.10$^{+0.56}_{-0.95}$ & 21.1$^{+0.01}_{-0.02}$  & 19,2 \\
iPTF16fnl$^c$ & H+He$^*$     & 5.90$^{+0.15}_{-0.06}$ & 0.98$^{+0.03}_{-0.87}$ & 0.88$^{+0.14}_{-0.03}$ & -3.8$^{+1.26}_{-0.10}$ & 0.84$^{+0.09}_{-0.08}$ & 0.88$^{+0.42}_{-0.04}$ & -1.2$^{+1.00}_{-1.06}$ & 20.2$^{+0.34}_{-0.32}$  & 20,2 \\
\hline
\end{tabular}
\\
\begin{flushleft}
Data sources: 1: \citet{Holoien2014}, 2: \citet{Hinkle2021a}, 3: \citet{Holoien2016a}, 4:  \citet{Holoien2016}, 5: \citet{Nicholl2019}, 6: \citet{Leloudas2019}, 7: \citet{Holoien2020}, 8: \citet{vanVelzen2021a}, 9: \citet{Gomez2020}, 10: \citet{Holoien2019}, 11: \citet{Cannizzaro2021a}, 12: \citet{Nicholl2020}, 13: \citet{Gezari2008}, 14: \citet{Wyrzykowski2017}, 15: \citet{Gezari2012}, 16: \citet{Chornock2014}, 17: \citet{Arcavi2014}, 18: \citet{vanVelzen2011}, 19: \citet{Hung2017}, 20: \citet{Blagorodnova2017}. \\
$^a$ Spectroscopic class according to \citet{vanVelzen2021}. A $*$ indicates a detection in X-rays;
$^b$ Exclude late-time data (see section \ref{sec:data});
$^c$ Shortened prior on $t_{\rm exp} < 30$\,d.;
$^d$ Although listed as TDE-H by \citet{vanVelzen2021}, we classify here as TDE-H+He due to the detection of He\,II and likely N\,III by \citet{Short2020}. We have verified that the re-classification of this event does not change our statistical conclusions.
\end{flushleft}
\label{tab:tdes}
\end{table*}

The physics of the model is described in detail by \citet{Guillochon2014} and \citet{Mockler2019}, but we summarise some key details here. Mass fallback curves, $\dot{m}_{\rm GR13}(t)$, were derived by \citet{Guillochon2013} from simulated disruptions of 1\,\M\ polytropic stars around a $10^6$\,\M\ SMBH, with a range of impact parameters. {The impact parameter, $\beta$, is the ratio of the tidal radius, $R_{\rm t}=(M_\bullet/M_*)^{1/3}R_*$, to the orbital pericentre, $R_{\rm p}$}. Analytic relations are used to scale the fallback curves to account for different black hole masses and stellar properties: 
\begin{equation}\label{eq:scale}
    \dot{m} = \dot{m}_{\rm GR13}(\beta) \left(\frac{M_\bullet}{10^6 {\rm M}_\odot}\right)^{-1/2} \left(\frac{M_*}{{\rm M}_\odot}\right)^{2} \left(\frac{R_*}{{\rm R}_\odot}\right)^{-3/2},
\end{equation}
where the stellar radius $R_*$ is defined for a star of a given mass using relations from \citet{Tout1996}, and $\dot{m}_{\rm GR13}$ is interpolated numerically for a desired $\beta$. {Because the degree of disruption at a given impact parameter depends on the density structure of the star \citep{Guillochon2013}, \textsc{mosfit} works in terms of a `scaled' impact parameter, $b$, which is defined such that $b\ge1$ corresponds to a complete disruption and $b=0$ to no disruption, regardless of the mass and radius of the star \citep[for further discussion see][]{Mockler2019}.}

To produce an observable TDE, some fraction of the energy from fallback must be converted into radiation. The common assumption is 
\begin{equation}
    \dot{E}_{\rm rad} =\epsilon \dot{m}c^2,
\end{equation}
where the efficiency $\epsilon$ allows us to remain agnostic about the energy conversion process. The theoretical range is from $\lesssim 10^{-2}$ for stream-stream collisions \citep{Jiang2016} up to 0.4 for accretion onto a maximally rotating SMBH. If conversion is prompt, the light curve will follow the shape of the fallback curve. In reality, this conversion will take a finite time, captured in the model by the inclusion of the viscous delay time, giving a bolometric luminosity
\begin{equation}\label{eq:bol}
    L(t) = \frac{1}{t_\nu} e^{-t/t_{\nu}} \int_0^t \epsilon \dot{m}(t') c^2 e^{t'/t_{\nu}}dt'.
\end{equation}
We note the similarity of this expression to the \citet{Arnett1982} solution for supernova light curves. For this solution to hold strictly, a key assumption of the \textsc{mosfit} TDE model is that both $\epsilon$ and $t_{\nu}$ are time-independent.

Rather than fitting bolometrically, \textsc{mosfit} uses multicolour information by fitting the broadband photometry directly. This requires a model for the spectral energy distribution. The colours of TDEs suggest that radiation is reprocessed by an extended photosphere, rather than emanating directly from a compact accretion disk. We relate the photospheric radius {to the Eddington ratio via the power-law \citep{Guillochon2014,Mockler2019}:
\begin{equation}\label{eq:phot}
    R_{\rm ph}(t) = R_{\rm ph,0} {a_{\rm p}} (L(t)/L_{\rm Edd})^{l_{\rm ph}},
\end{equation}
where
\begin{equation}
    a_p = \left(8 G M_\bullet \left(\frac{t_{\rm peak}-t_0}{\pi}\right)^2\right)^{1/3}
\end{equation}
is the semi-major axis of the material at peak fall-back rate.} This mimics a reprocessing layer that can expand or contract with the instantaneous luminosity. {As discussed by \citet{Mockler2019}, this allows us to remain agnostic about the heating mechanism (accretion or collisions), but does assume that all radiation is thermalised at the scale of the photosphere. This photosphere can form from disk-driven outflows \citep{Metzger2016,Nicholl2020,Parkinson2022} in the accretion-powered scenario, or shocks \citep{Lu2020} or unbound debris streams in a collision-powered scenario \citep{Jiang2016}. Our prior allows $R_{\rm ph}(t)$ to vary from the scale of the innermost stable circular orbit, appropriate for a very compact disk and atmosphere, up to 100 times the semi-major axis of the bound debris for typical Eddington ratios.}

The emitted spectrum is assumed to be a blackbody, and the temperature is determined by the instantaneous luminosity and radius. Although TDE spectra are regularly approximated as a thermal continuum, our assumption of a simple blackbody photosphere is another important caveat to the model.

\textsc{mosfit} assumes a Planck cosmology \citep{Planck2016} to calculate the distance to each TDE. We use the same priors in this study as in \citet{Mockler2019}, listed in Table \ref{tab:priors}, except {for the efficiency, which we broaden following \citet{Mockler2021}}. For some TDEs, we also shorten the prior on $t_0$, when the rise is poorly resolved -- such cases are noted in Table \ref{tab:tdes}. The posterior distributions of the model parameters are evaluated using dynamic nested sampling with \textsc{dynesty} \citep{Speagle2020}. Our likelihood function is:
\begin{equation}
    \ln\mathcal{L}=-\frac{1}{2}\sum_{i=1}^{n}\left[\frac{(O_i-M_i)^2}{\sigma_i^2+\sigma^2}+\ln(\sigma_i^2 + \sigma^2)\right],
\end{equation}
where $O_i$ and $M_i$ are the set of observed and model magnitudes, and $\sigma_i$ are the observational uncertainties. All fits are run to convergence on the University of Birmingham BlueBEAR computing cluster.

\begin{figure*}
\centering
	\includegraphics[width=0.33\textwidth]{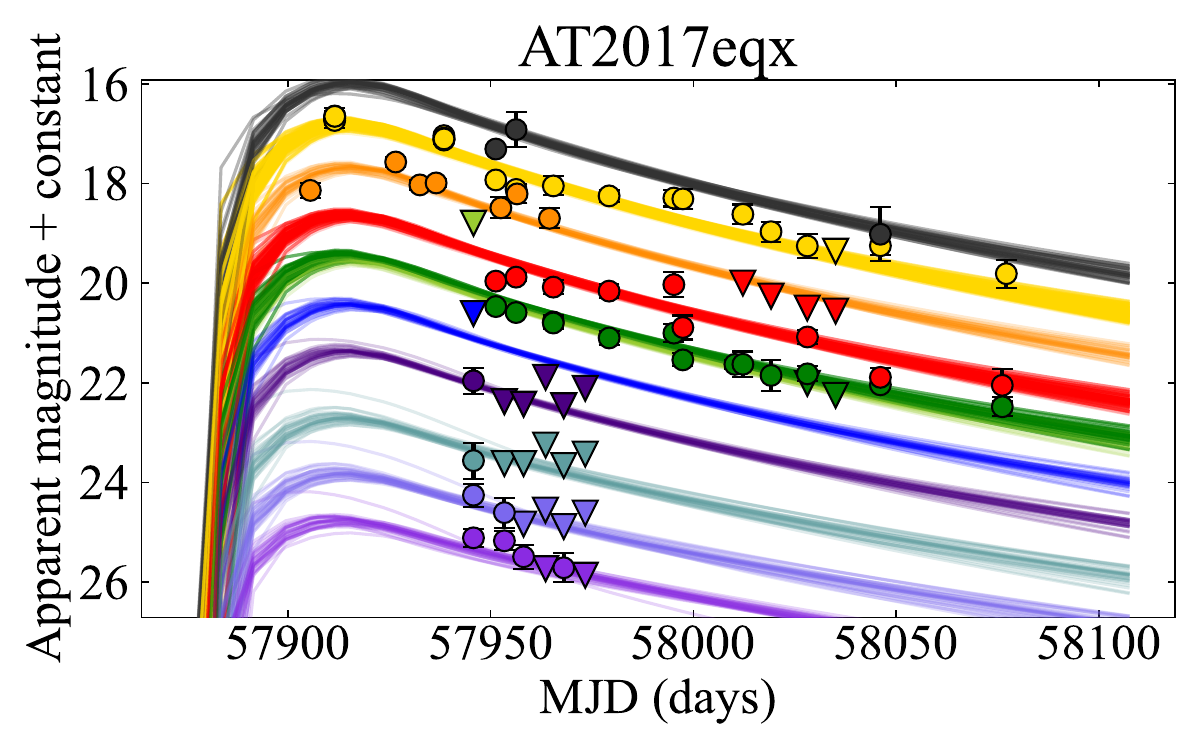}
	\includegraphics[width=0.33\textwidth]{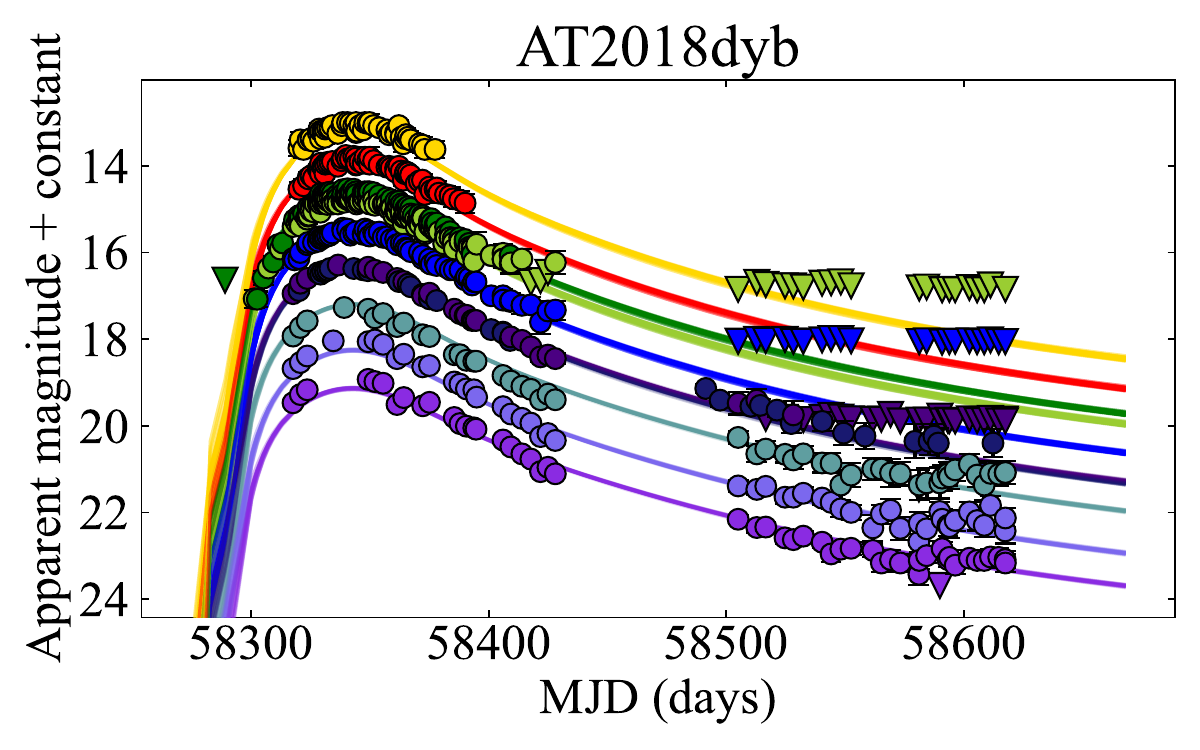}
	\includegraphics[width=0.33\textwidth]{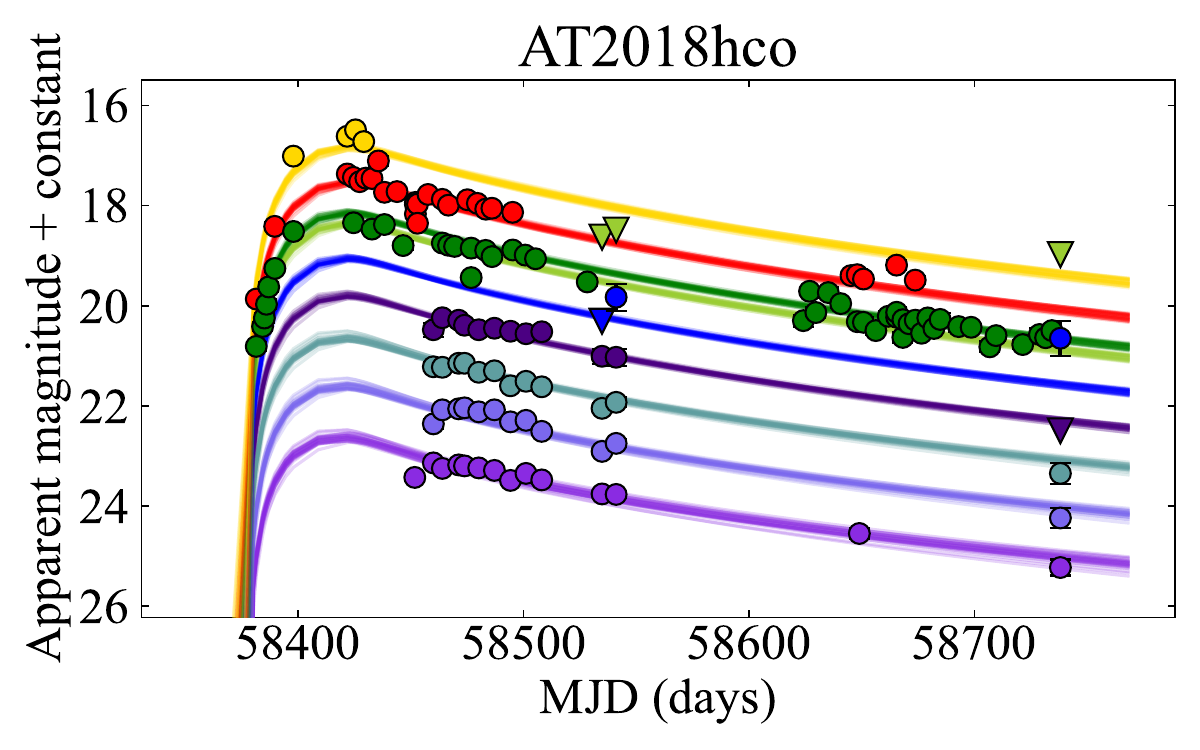}
	\includegraphics[width=0.33\textwidth]{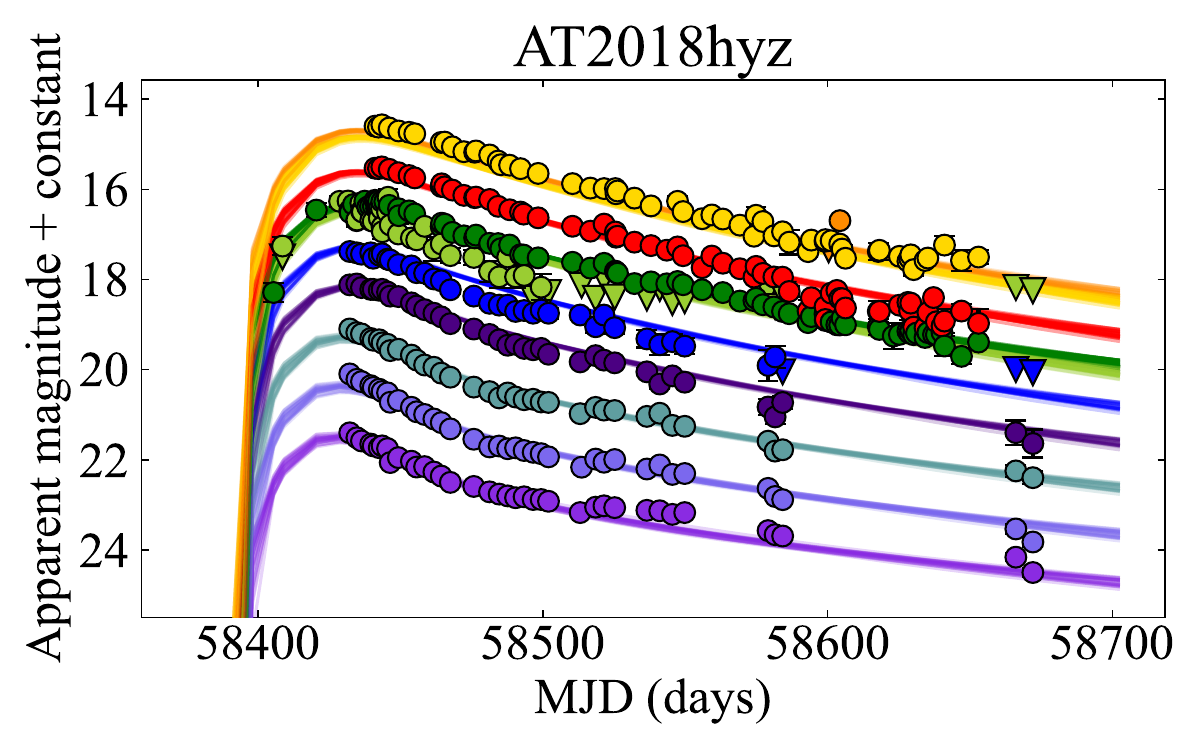}
	\includegraphics[width=0.33\textwidth]{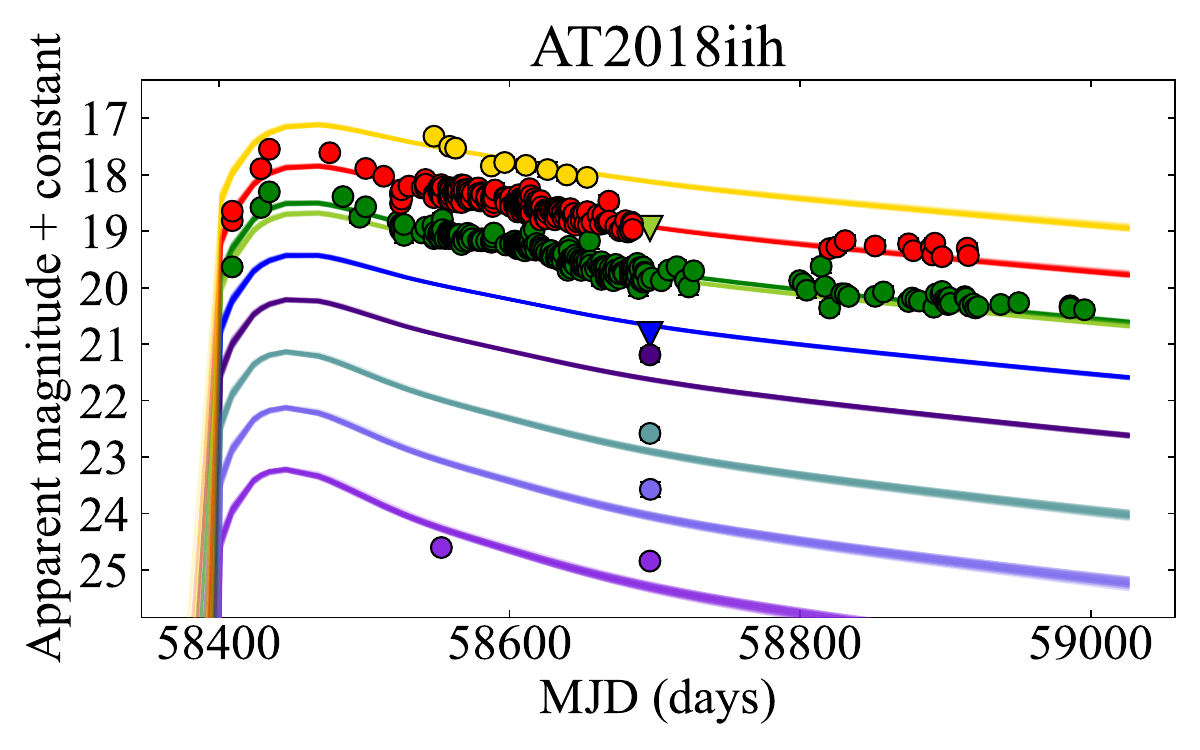}
	\includegraphics[width=0.33\textwidth]{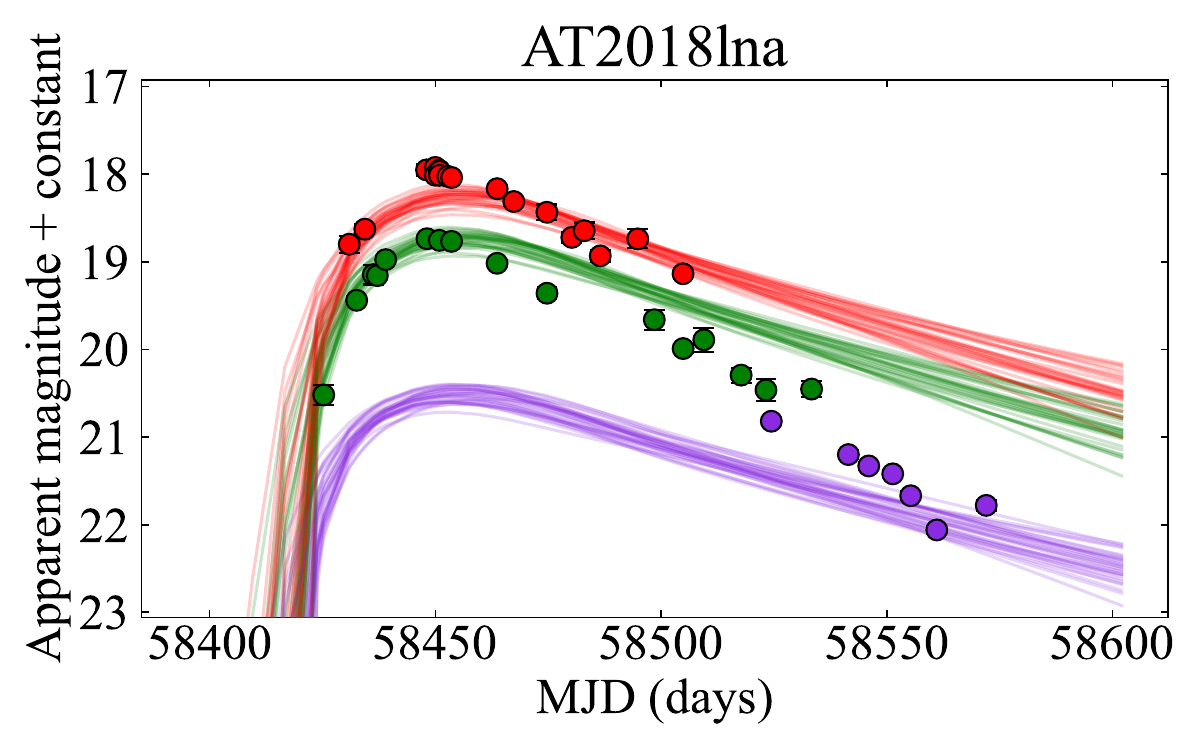}
	\includegraphics[width=0.33\textwidth]{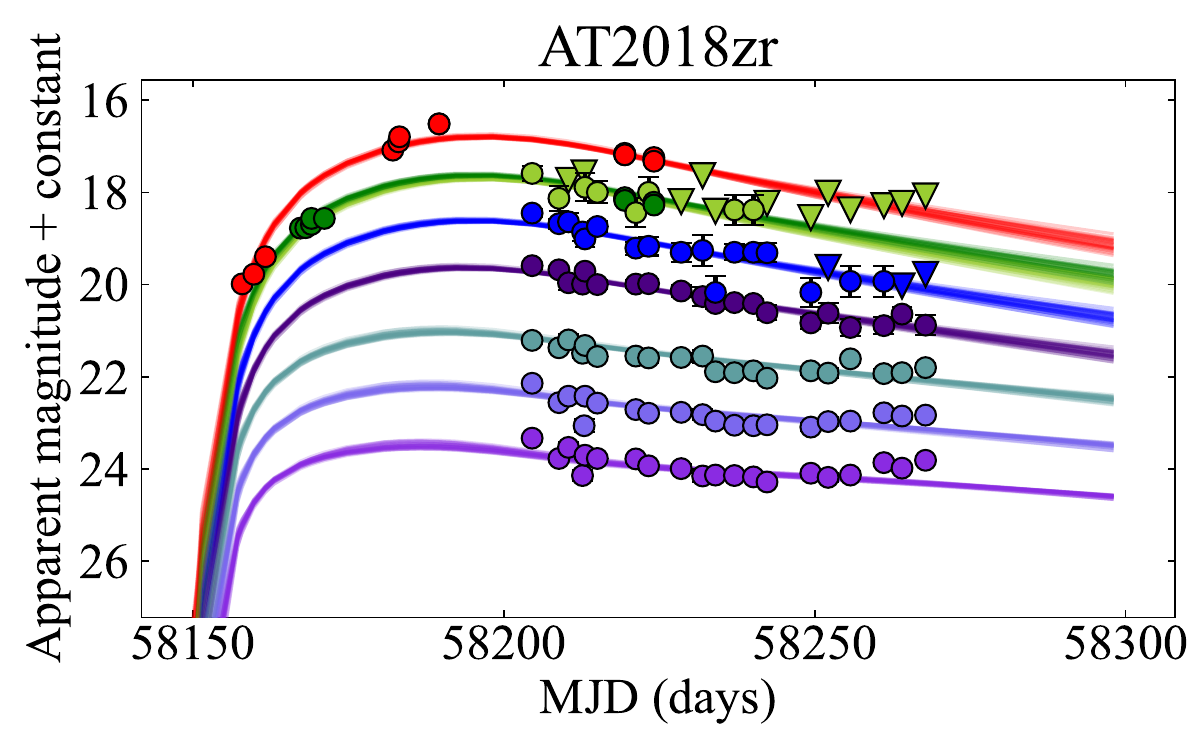}
	\includegraphics[width=0.33\textwidth]{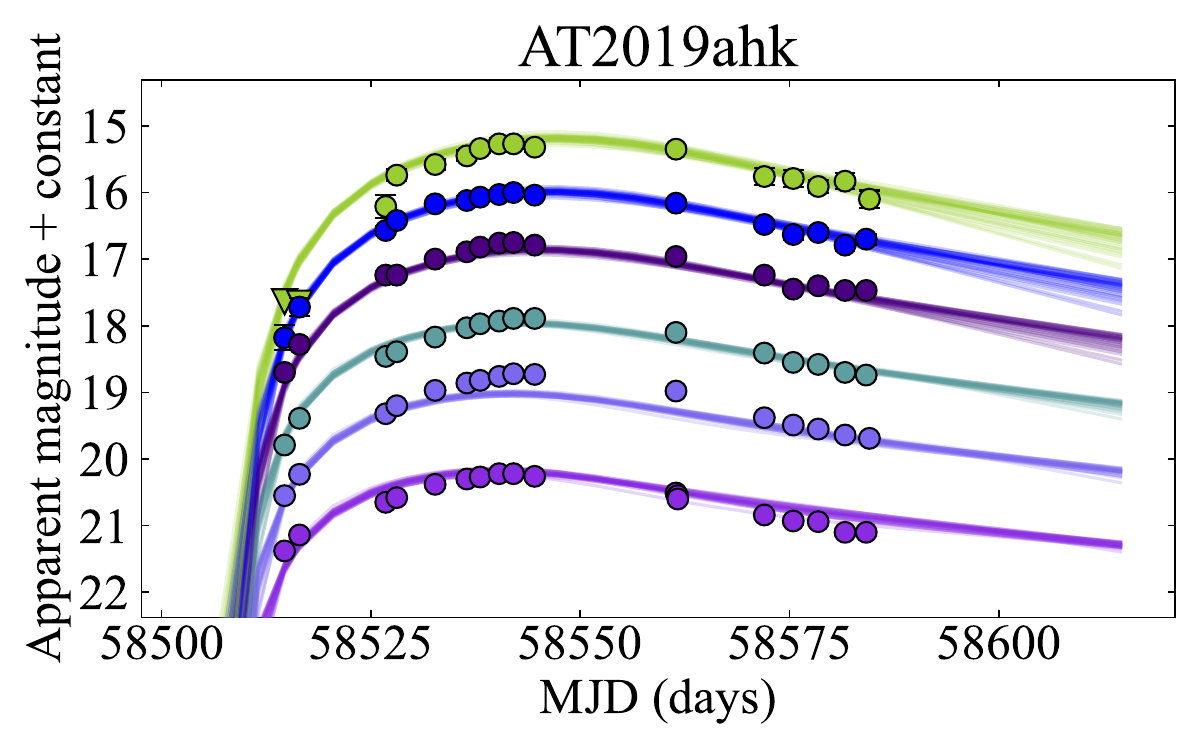}
	\includegraphics[width=0.33\textwidth]{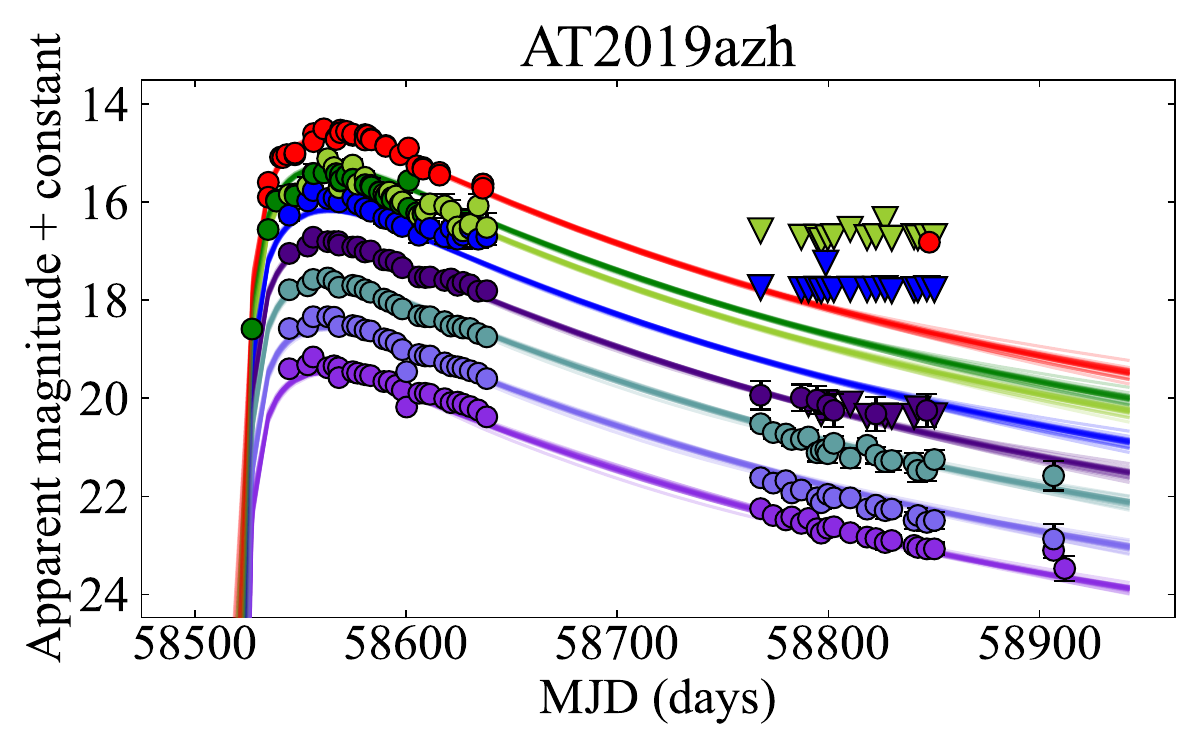}
	\includegraphics[width=0.33\textwidth]{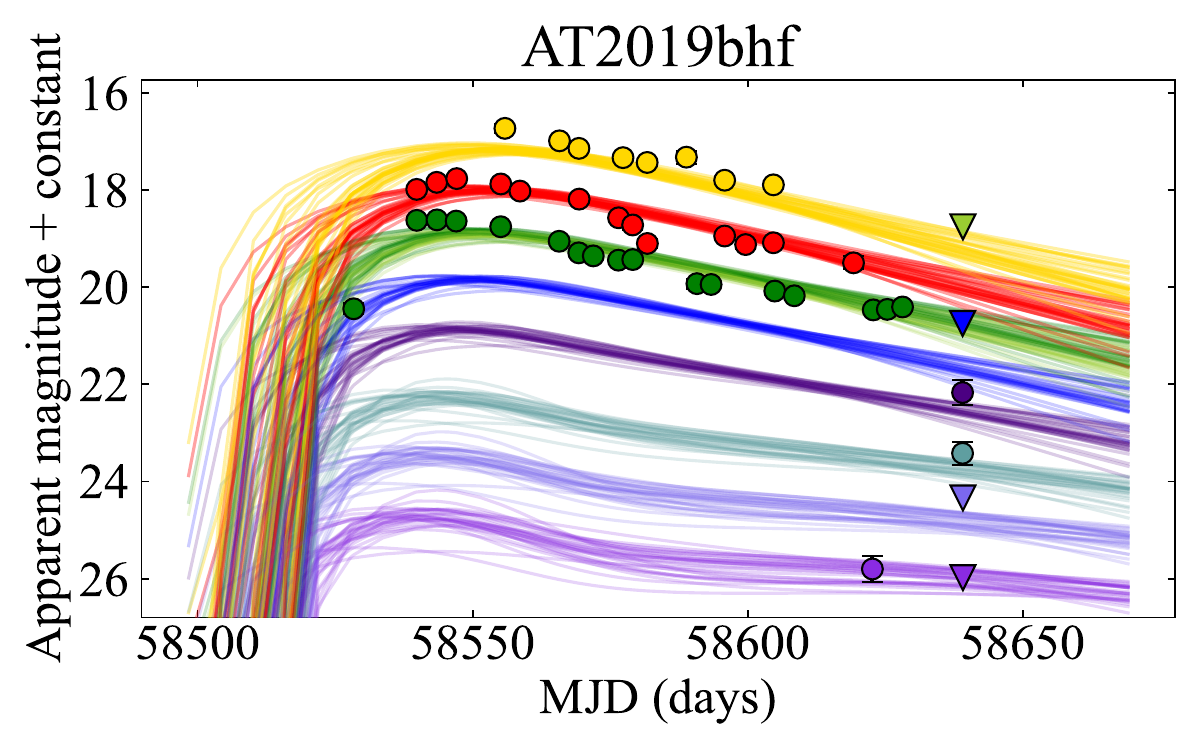}
	\includegraphics[width=0.33\textwidth]{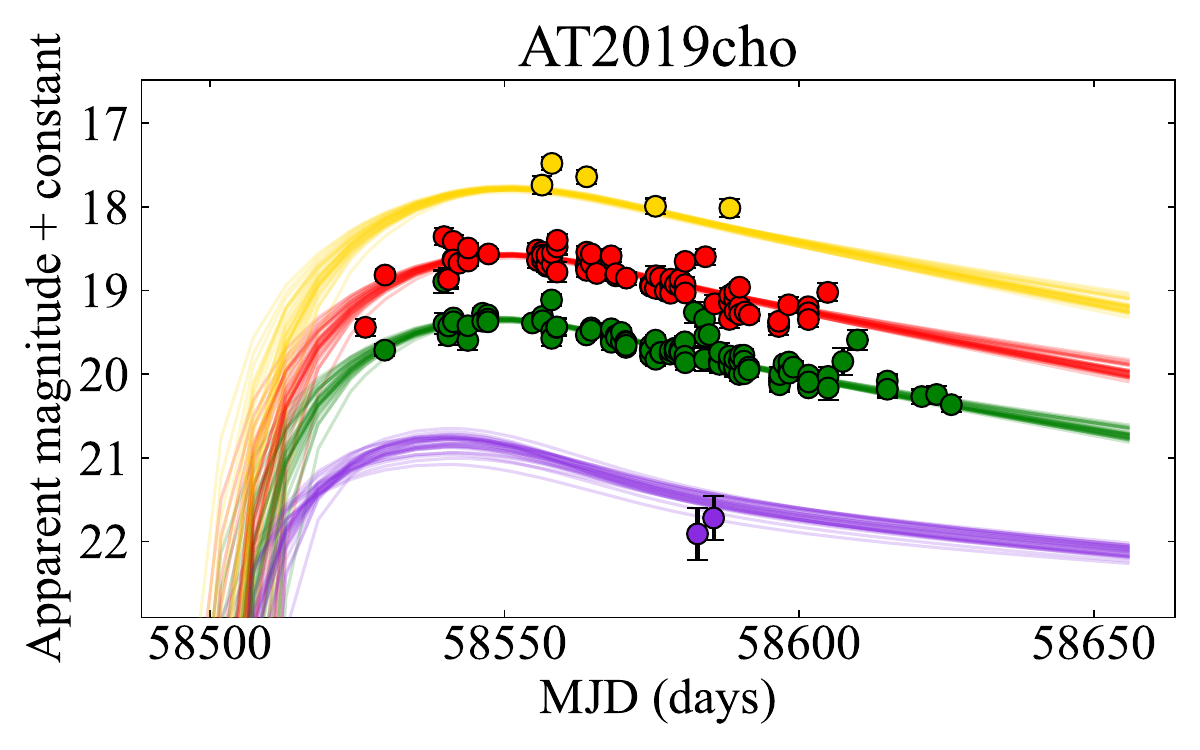}
	\includegraphics[width=0.33\textwidth]{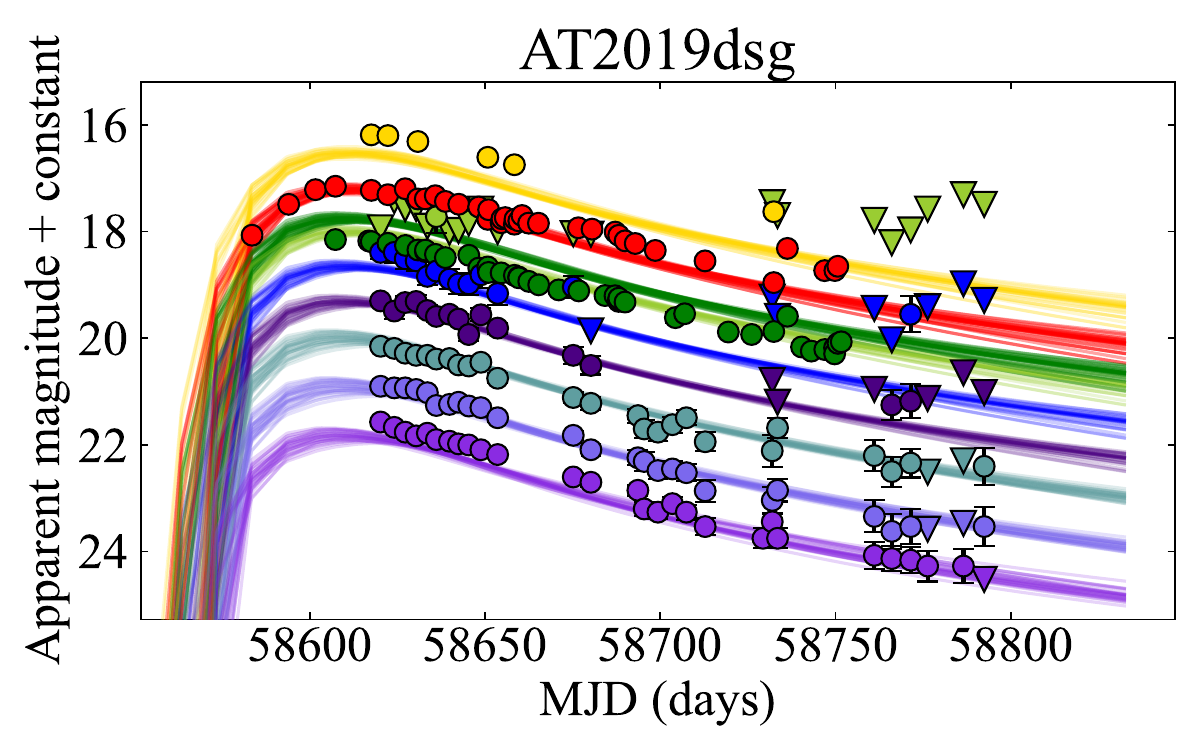}
	\includegraphics[width=0.33\textwidth]{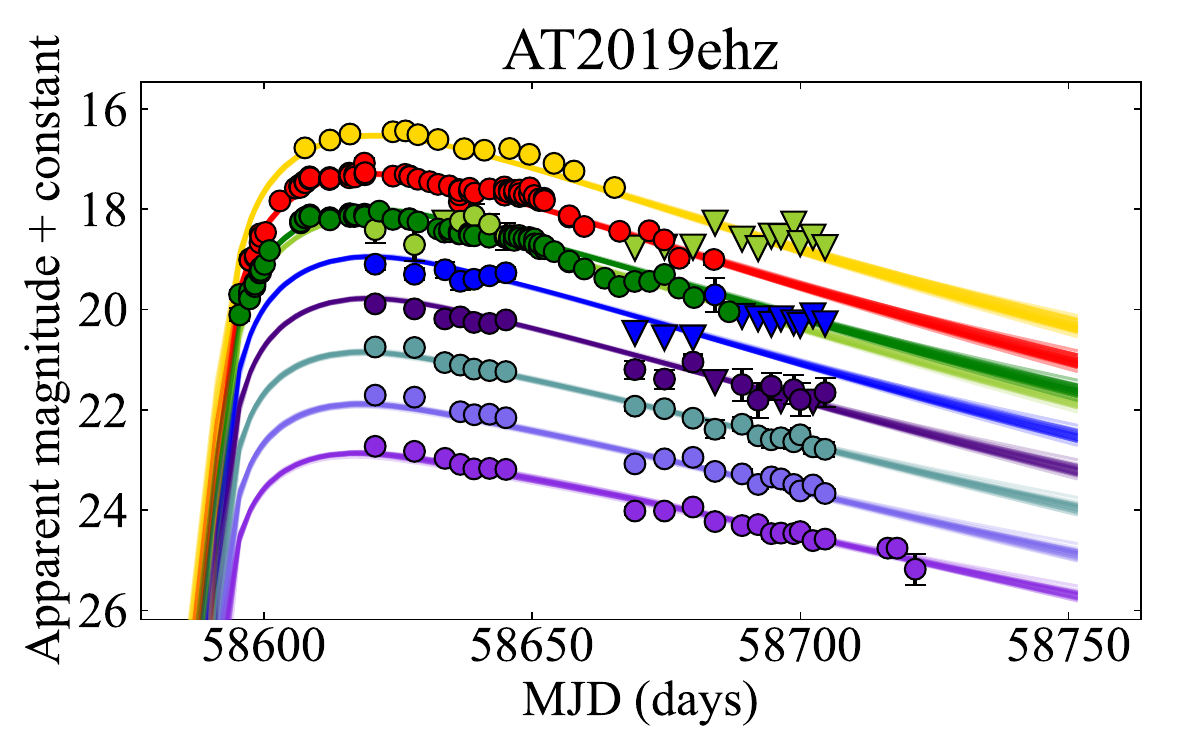}
	\includegraphics[width=0.33\textwidth]{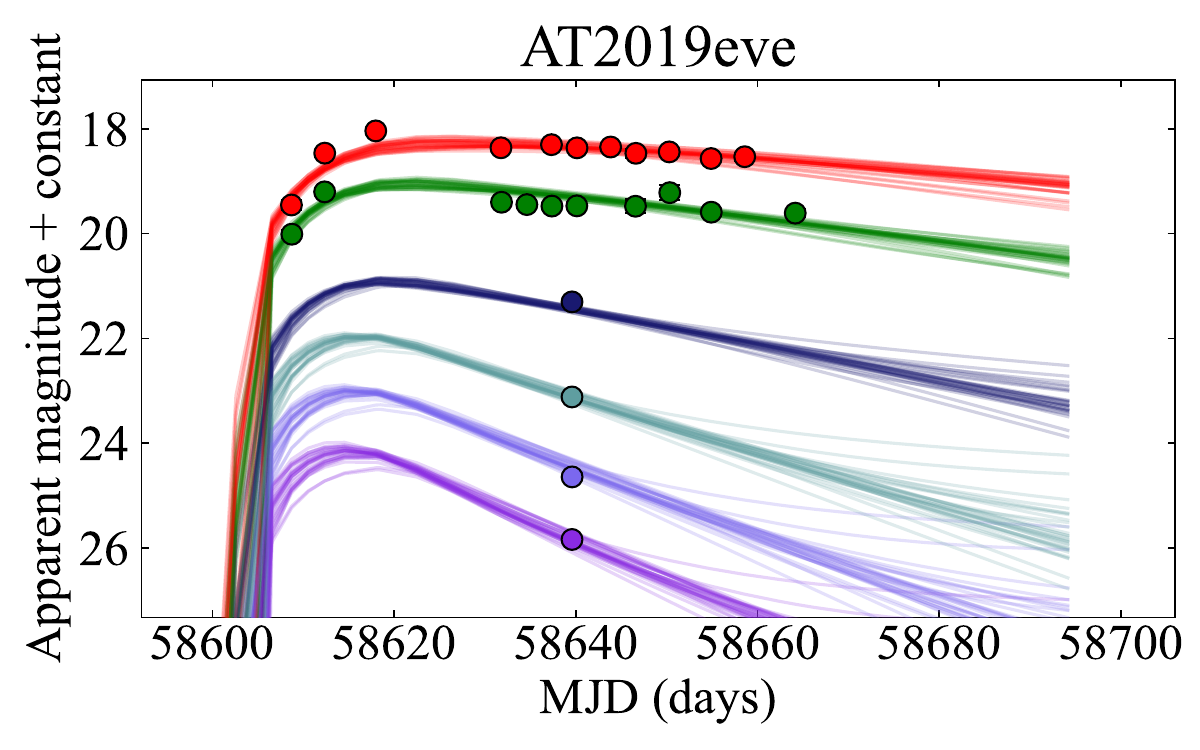}
	\includegraphics[width=0.33\textwidth]{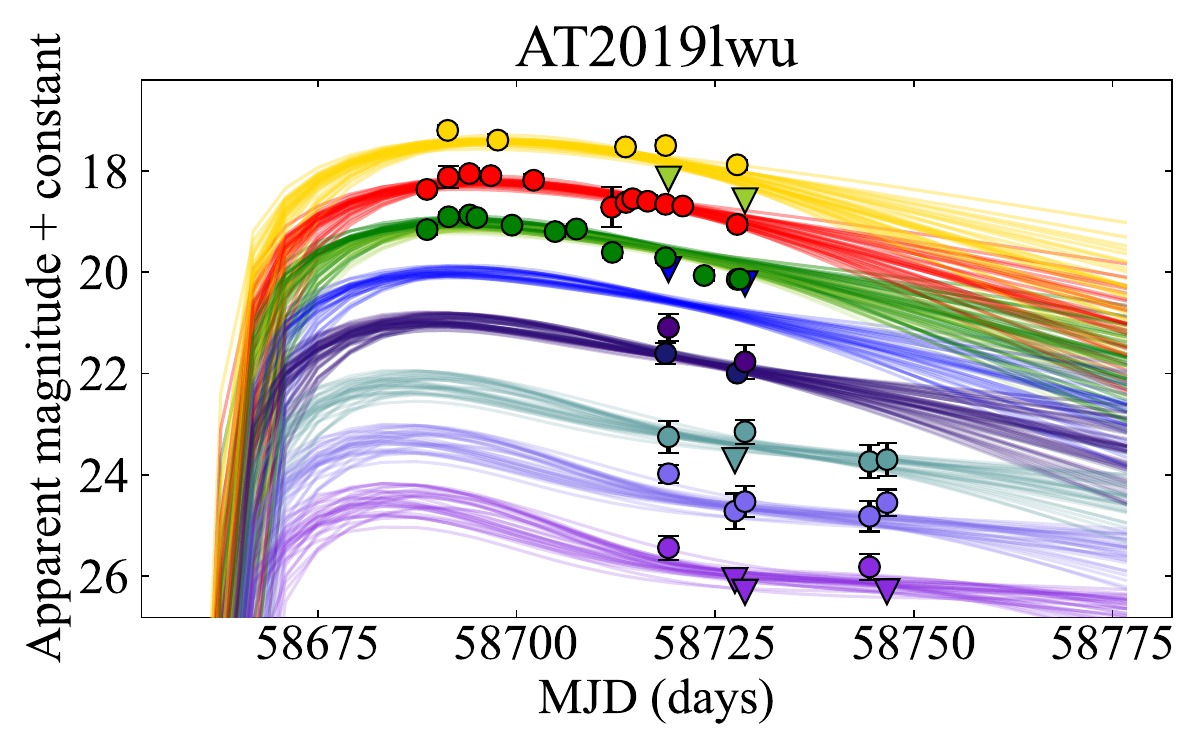}
	\includegraphics[width=0.33\textwidth]{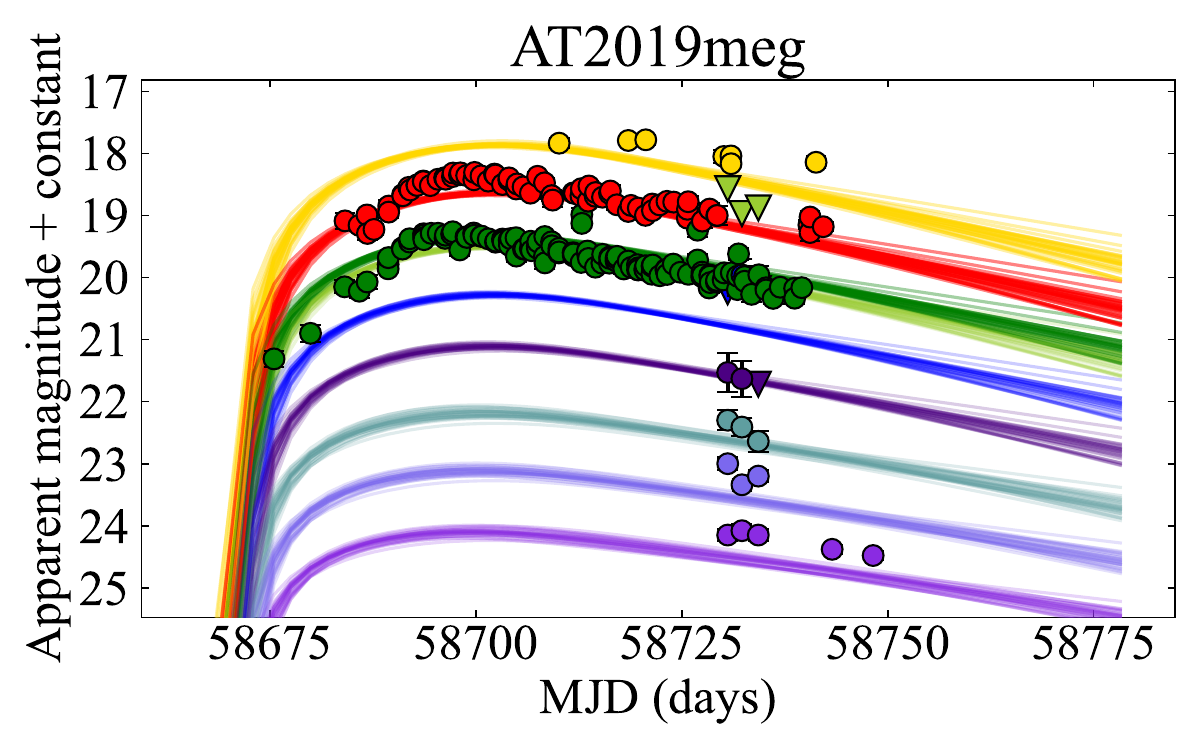}
	\includegraphics[width=0.33\textwidth]{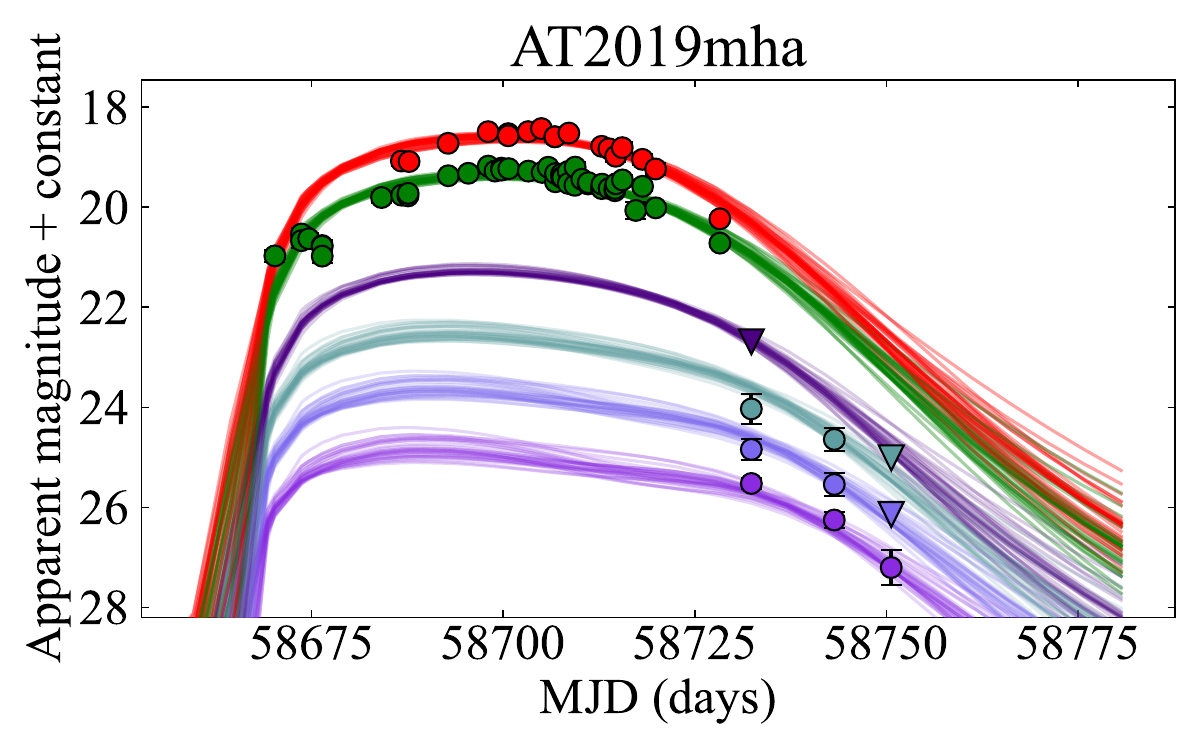}
	\includegraphics[width=0.33\textwidth]{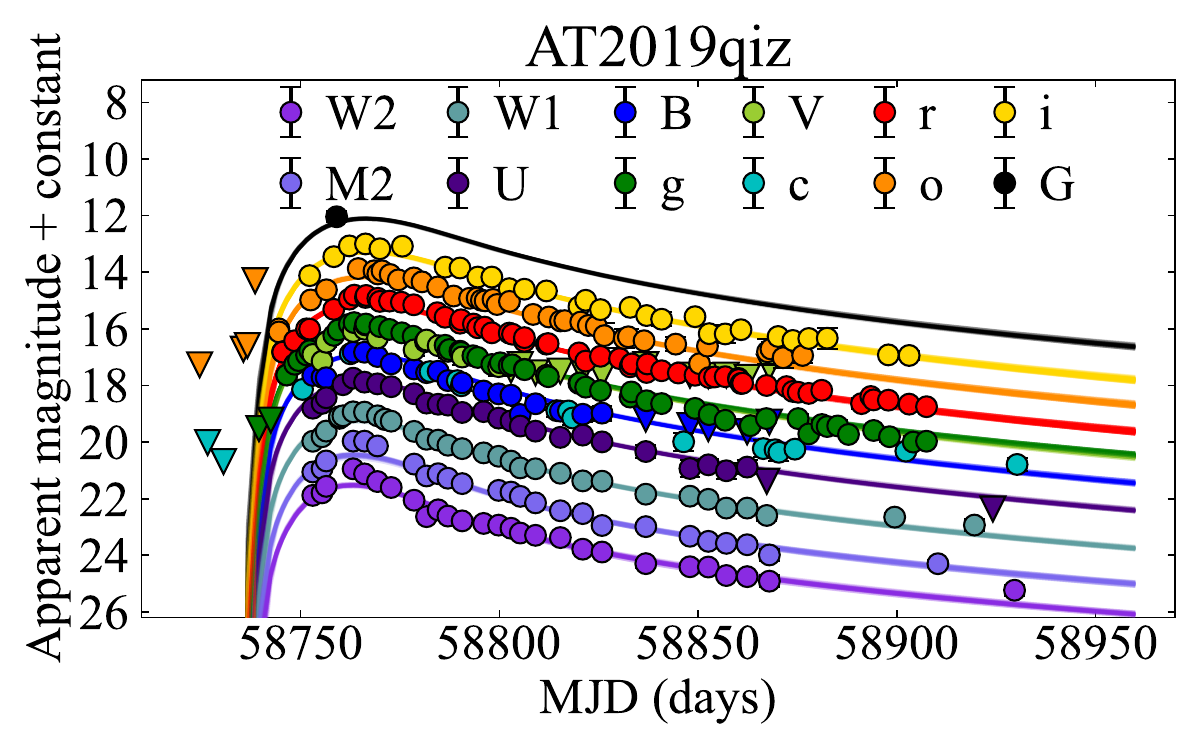}
    \caption{Fits to all TDEs in the sample discovered from 2017 onwards (earlier events were included in the \textsc{mosfit} study by \citealt{Mockler2019}, and fits can be seen in that work). In each panel the $g$ and/or $V$ bands show the true apparent magnitude, and every other band is progressively offset by an additional magnitude up (redder band) or down (bluer band) for visibility.}
    \label{fig:lcs}
\end{figure*}

\begin{table*}
\caption{Results of two-sample Kolmogorov-Smirnov tests comparing the distributions of parameter values between different TDE spectroscopic classes. The value $p_{\rm null}$ is the probability of the null hypothesis ("parameter values for both classes are drawn from the same distribution") given the data. Differences that are significant at the 5\% level ($p_{\rm null}<0.05$) are highlighted in \textbf{bold font}, those that are significant at the 1\% level ($p_{\rm null}<0.01$) are \emph{\textbf{bold and italic}}. `Other' refers to the unclassified TDEs from Table \ref{tab:tdes}.
}

\begin{tabular}{c|cccccc}
Parameter  & \multicolumn{6}{c}{$p_{\rm null}$ for two-sample KS test (number of events)} \\
\hline
  & H vs H+He & H vs He & H vs He/H+He & He vs H+He & H vs Other & He/H+He vs Other \\
 & (12 vs 11) & (12 vs 4) & (12 vs 15) & (11 vs 4) & (12 vs 5) & (15 vs 5) \\
\hline
$M_{\bullet}$ & 0.225 & 0.112 & 0.126 & 0.408 & 0.215 & 0.934 \\
$M_{*}$ & 0.810 & 0.991 & 0.839 & 0.824 & 0.463 & 0.770 \\
$b$ & \textbf{0.012} & \textit{\textbf{0.005}} & \textit{\textbf{0.001}} & \textbf{0.022} & \textit{\textbf{0.002}} & 0.343 \\
$\epsilon$ & 0.694 & 0.404 & 0.839 & 0.144 & 0.463 & 0.549 \\
$R_{\rm ph,0}$ & \textit{\textbf{0.004}} & 0.859 & \textbf{0.030} & 0.329 & \textbf{0.021} & 0.343 \\
$l_{\rm ph}$ & 0.225 & 0.991 & 0.434 & 0.824 & \textit{\textbf{0.007}} & 0.052 \\
$t_{\nu}$ & 0.108 & 0.641 & 0.151 & 0.895 & 0.930 & 0.205 \\
$n_{H}$ & \textbf{0.039} & 0.215 & 0.268 & 0.063 & 0.328 & 0.205 \\

\hline
\end{tabular}
\label{tab:ks}
\end{table*}

\subsection{TDE sample}
\label{sec:data}

We select our TDE sample from the compilation in \citet{vanVelzen2021a}. This consists of 17 events discovered by the Zwicky Transient Facility \citep[ZTF;][]{Bellm2019} and 22 earlier events from the literature. Of these 39 events, we omit four that have data insufficient for detailed light curve modelling (GALEX D23H-1, PTF09axc, AT2018bsi), and further exclude ASASSN-15lh due to its so-far unique properties and debated nature \citep{Dong2016,Leloudas2016,Margutti2017}. Finally, {we were unable to find} a satisfactory fit for AT2018fyk \citep[which has a double-peaked light curve;][]{Wevers2019} or iPTF15af \citep[{which has a relatively flat optical light curve;}][]{Blagorodnova2019}. The remaining 32 events consist of 12 TDE-H, 11 TDE-H+He, 4 TDE-He, and 5 unknown or unclassified events.

Of our 32 TDEs, 14 were previously analysed using \textsc{mosfit} by \citet{Mockler2019}. However, a substantial fraction of TDE imaging data is from the UV-Optical Telescope (UVOT) on the Neil Gehrels \textit{Swift} Observatory, and in November 2020 an issue was identified with the calibration files used to reduce recent UVOT data. We therefore reanalyse these events after replacing all UVOT data with the corrected photometry provided by \citet{Hinkle2021a}. Additionally, we have swapped out the \textsc{emcee} sampler \citep{ForemanMackey2013}, used in previous \textsc{mosfit} studies, for \textsc{dynesty}, which may be more efficient at exploring complex posteriors in high-dimensional spaces.

For most events, we include all of the publicly available data. However, a few events have light curves extending over several years. \citet{vanVelzen2019} identify an additional luminosity component from a long-lived, slowly evolving disk in several TDEs with UV observations at these phases. This component is not included in the fallback light curves produced by \textsc{mosfit}. We therefore exclude data points later than MJD 57165 for ASASSN-14li, 53600 for GALEX D3-13, and 56000 for PS1-10jh, where \citet{vanVelzen2019} show the stable disk component to dominate.

Table \ref{tab:tdes} lists the full TDE sample and their spectroscopic types, the sources of data used, and any additional constraints used in the model fits.

\section{Results}
\label{sec:results}

In this section we discuss the model posteriors and the main statistical results derived from our light curve fits. The complete set of light curves and model parameters are provided in Figure \ref{fig:lcs} and Table \ref{tab:tdes}.

\subsection{Inferred physical parameters}

The SMBH masses of observed TDEs lie in the range $10^{5.8}-10^{7.2}$\,\M, with a mean of $10^{6.6}$\,\M\ and nearly all TDEs falling between $10^6-10^7$\,\M. The disruptions are all consistent with low-mass stars of $0.08-1.3$\,\M, with the exception of AT2019bhf (2.6\,\M). Most of the stars have masses $\lesssim 0.5$\,\M. The scaled impact parameter $b$ ranges from $0.3-1.2$, with a mean of 0.82. The efficiency spans more than two orders of magnitude, from $\approx 10^{-3}$ to $\approx 0.4$, and appears to correlate with SMBH mass (section \ref{sec:disc}). The photospheric parameters in equation \ref{eq:phot} also span several orders of magnitude and are correlated with each other. The final two parameters are the viscous timescale and host galaxy column density. While these span a wide range, few TDEs in the observed sample require long viscous delays or high extinction.

We note that for some TDEs, we find differences in the inferred parameters compared to \citet{Mockler2019}. {This can be traced to broadening the radiative efficiency prior, the changes in UVOT calibration for some events, and the ability of the nested sampler to resolve multi-modal posteriors. The differences primarily affect the posteriors for star mass and efficiency. We discuss this in more detail in Appendix \ref{sec:caveats}.}

\begin{figure*}
\centering
	\includegraphics[width=\textwidth]{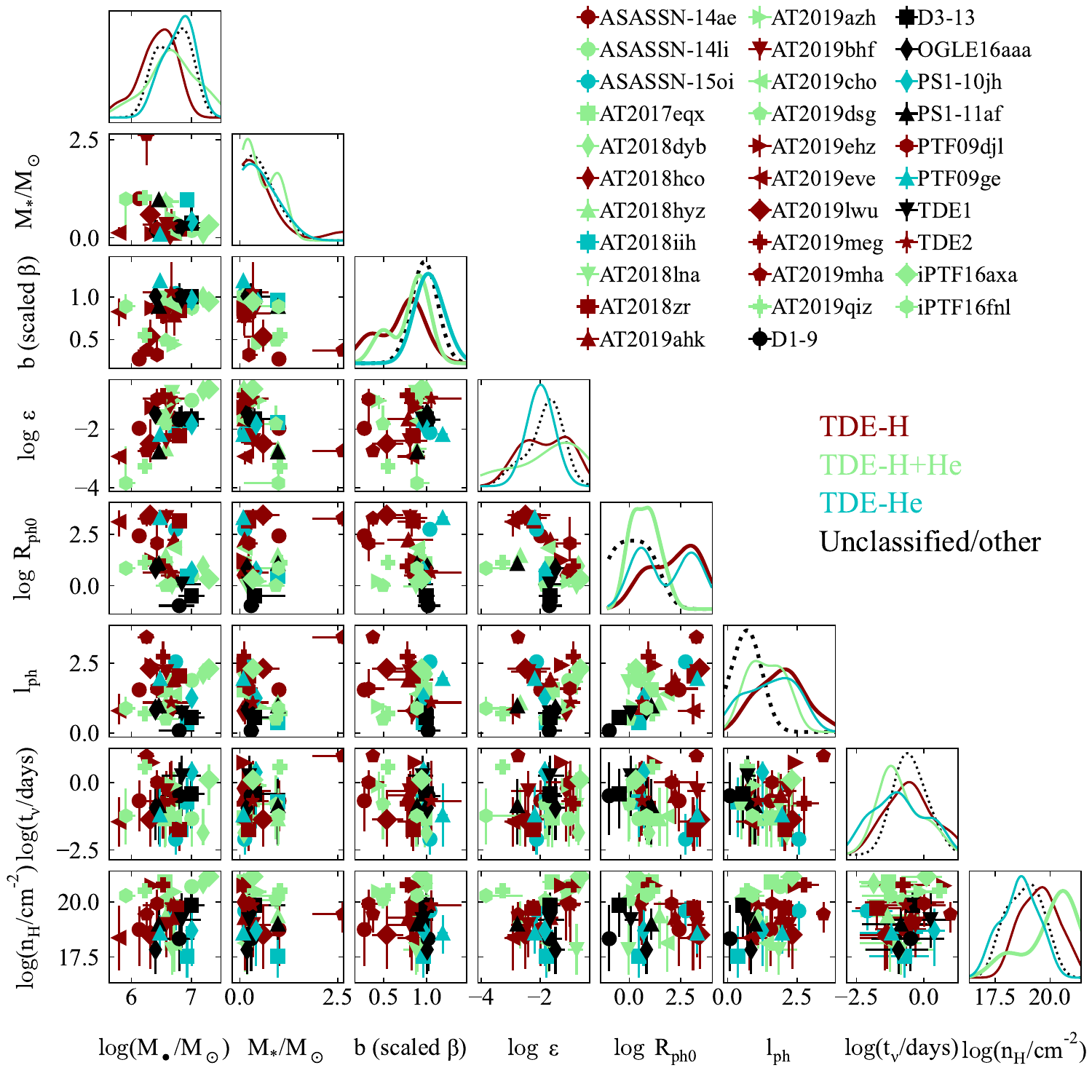}
    \caption{Corner plot showing the median and $1\sigma$ uncertainties of the model posteriors for physical parameters. Each point is coloured according to the spectroscopic classification of the relevant TDE. The diagonal shows Gaussian kernel density estimates for the underlying distribution in each class. We smooth with a bandwidth $h_{i,j}=\sigma_{i,j} n_j^{-1/5}$ for each parameter, where $\sigma_{i,j}$ is the standard deviation of parameter $i$ for spectroscopic class $j$ and $n_j$ the number of events in that class. We highlight with thicker lines those distributions that are statistically distinct from at least one other class, according to a KS test.}
    \label{fig:corner}
\end{figure*}

\subsection{Statistical differences between TDE classes}

Figure \ref{fig:corner} shows a corner plot of the best-fit TDE model parameters, with events coloured by spectroscopic type. Some trends are immediately visible, suggesting systematic differences in the derived parameters between TDE sub-classes. To quantify these differences, we apply a series of two-sample Kolmogorov-Smirnov (KS) tests for each parameter, between the various classes. We consider the He and H+He TDEs both as separate classes, and as a larger class of He/H+He (i.e., TDEs with high-excitation emission lines).

The full results of our KS tests are listed in Table \ref{tab:ks}. We highlight significant differences according to the usual criterion, $p_{\rm null}<0.05$, where $p_{\rm null}$ is the probability that both sets of parameter values are drawn from the same distribution. However, given that this Table includes 48 separate tests, the probability from a binomial distribution of obtaining at least one success by chance is 90\% (though the probability of obtaining 10 successes, as found here, is $10^{-4}$). We therefore stress that several tests have KS statistics with $p_{\rm null}<0.01$; the probability of obtaining more than one such result by chance is only 8\%.

We find several significant differences in the parameters controlling the size of the photosphere, $R_{\rm ph,0}$ and $l_{\rm ph}$. TDE-H and TDE-H+He events have different $R_{\rm ph,0}$, with $p_{\rm null}=0.004$. Differences in photospheric radii between H and H+He TDEs were also identified by \citet{vanVelzen2021a}, \citet{Hinkle2020} and \citet{Charalampopoulos2021}, in the sense that TDE-H have larger radii than TDE-H+He, in agreement with our results. 
Significant differences are also found between the Other group and all classified TDEs. The Other events have the smallest photospheric radii, and the weakest scaling with the Eddington ratio. We caution however that this group is small in our sample (5 events), and may not be a homogeneous class.

Our more novel finding is that the $b$ parameter differs between the TDE-H class and every other group. This is a highly significant result, with $p_{\rm null}<0.01$ for most of the relevant tests. This parameter is defined such that the fraction of stellar mass surviving as a bound remnant after disruption is zero for $b=1$, regardless of polytropic index (i.e. stellar density profile). Smaller values of $b$ correspond to partial disruptions, with some fraction of the star remaining bound, and larger values correspond to full disruptions of increasing penetration into the SMBH potential. The mean value for the TDE-H class is $\langle b\rangle=0.68$, whereas the TDE-H+He class has $\langle b\rangle=0.81$ and the TDE-He class has $\langle b\rangle=1.03$, i.e. a larger fraction of the disrupted mass is bound to the SMBH in the TDE-H+He and especially TDE-He events. {This also reflects differences in how close the stars encountered their SMBHs. Expressed in terms of $\beta=R_{\rm t}/R_{\rm p}$, we find $\langle \beta\rangle=0.89$, $\langle \beta\rangle=1.08$ and the $\langle \beta\rangle=1.24$ for the TDE-H, H+He and He subclasses, respectively.
}

We also find that TDE-H events may arise on average from less massive SMBHs, with $\langle \log(M_\bullet/M_\odot) \rangle = 6.4$, than the TDE-He/H+He events, which have $\langle \log(M_\bullet/M_\odot) \rangle = 6.8$. In this case the KS test results are not statistically significant. However, we note that none of the seven most massive SMBHs in the sample are associated with TDE-H events, and the probability of finding this by chance in a sample of this size is $\ll 1$\%.
Finally, TDE-H+He events in our sample seem to have larger extinctions (host galaxy column densities) than do TDE-H ($p_{\rm null}=0.04$).

\section{Discussion}
\label{sec:disc}

\subsection{Disrupted star demographics}

\begin{figure}
	\includegraphics[width=\columnwidth]{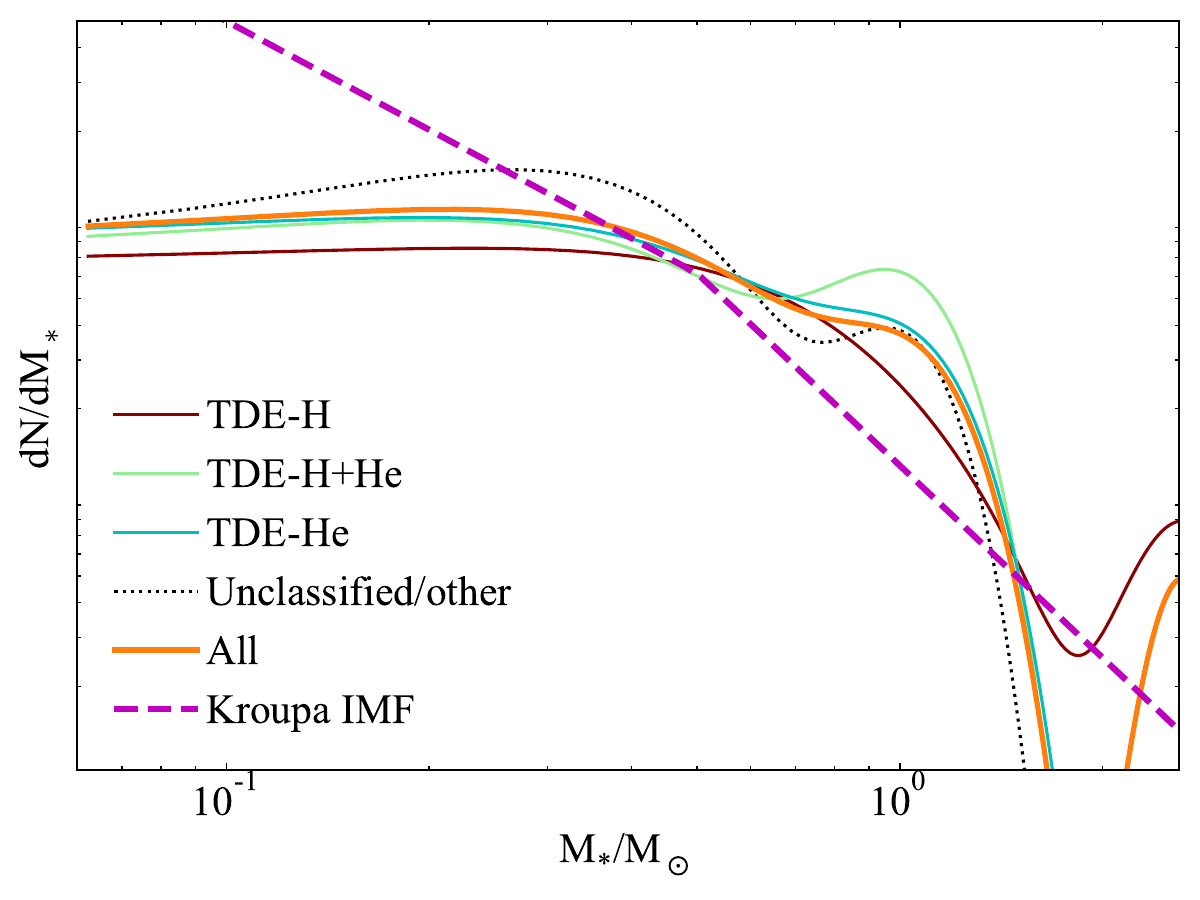}
    \caption{Kernel density estimates of the distribution of star masses. The distributions for each spectroscopic class of TDE are similar, and significantly flatter than a standard Kroupa initial mass function \citep{Kroupa2001}.}
    \label{fig:imf}
\end{figure}

The stellar masses returned by our fits favour low-mass stars in the range $\approx 0.1-1$\,\M, with similar distributions for all spectroscopic classes. In Figure \ref{fig:imf}, we compare the mass distribution for each TDE class, and for all TDEs combined, to a Kroupa initial mass function \citep[IMF;][]{Kroupa2001}. The Kroupa IMF consists of broken power laws, $dN/dM_*\propto M_*^{-\alpha}$, with $\alpha=0.3$ for $M_*<0.08$\,\M, $\alpha=1.3$ for $0.08<M_*<0.5$\,\M, and $\alpha=2.3$ for $M_*>0.5$\,\M. This IMF was used as the prior on the star mass in each fit (though note that this is not the same as using a heirarchical Kroupa prior on the population as a whole). 

We find that the distribution of the stellar mass posteriors is significantly flatter than the IMF. We quantify the discrepancy with respect to the IMF by computing the Kullback-Liebler divergence. We find $D_{\rm KL}=0.60$ between the combined TDE sample and the IMF. Although there is no formal threshold for a `significant' value in $D_{\rm KL}$, this is much larger than the divergence between the TDE-H and TDE-H+He mass distributions ($D_{\rm KL}=0.05$). 

Selection effects may play a role here. The fallback rate (determining the TDE luminosity) scales as $\dot{M}\propto M_*$ \citep{Ryu2020I,Law-Smith2020}, while the duration of this mass return scales as $t\propto M_*^{-1}$. Thus very low-mass stars, while more common, tend to produce faint and slowly-evolving TDEs, which may be harder to detect than sharper, brighter flares from more massive stars. 

However, it is likely that selection effects are not the only factor, and the \emph{intrinsic} mass function of TDEs is also flatter than a standard IMF. A star can be disrupted outside of the event horizon only by SMBHs below the Hills mass \citep{Hills1975}:
\begin{equation}
    M_\bullet \leq 9\times10^7 {\rm M}_\odot (R_*/{\rm R}_\odot)^{3/2} (M_*/{\rm M}_\odot)^{-1/2}.
\end{equation}
Therefore (because $R_*$ increases with $M_*$) the more massive the star, the larger the range of SMBH mass over which it can produce an observable flare, somewhat counteracting the relative scarcity of such stars in the standard IMF.

At very low masses, the distributions estimated with \textsc{mosfit} fall well below the predictions of the IMF, with no stars below $0.06$\,\M\ (the 90\% lower bound of the lowest-mass posterior), despite the prior favouring masses as low as 0.01\,\M. The upper end of the `missing' regime is close to the hydrogen-burning limit at $0.08$\,\M. Below this, brown dwarf stars cannot ignite core hydrogen fusion and are supported primarily by electron degeneracy pressure. This predicts a relatively flat mass-radius relation, with $R_*\sim 0.1$\,R$_\odot$ across the full mass range from $0.01-0.08$\,\M\ \citep[e.g.][]{Burrows1997,Chabrier2000,Burgasser2006}. The lack of brown dwarf TDEs inferred from our fits may therefore reflect the difficulty of disrupting stars that have average densities $>10$ times that of a solar-mass star.

\subsection{Efficiency and BH mass}
\label{sec:corr}

\begin{figure}
	\includegraphics[width=\columnwidth]{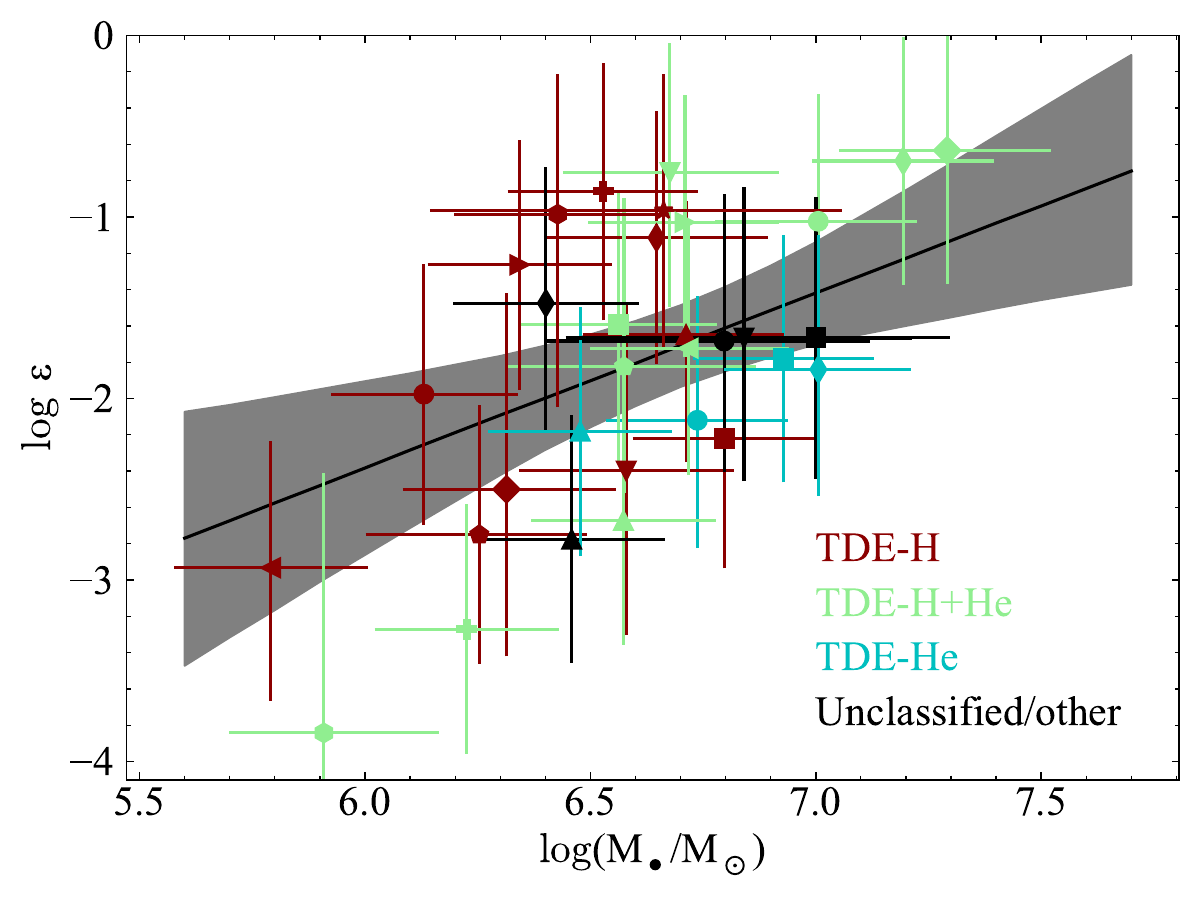}
    \caption{TDE radiative efficiency vs SMBH mass. Error bars include statistical and systematic errors \citep{Mockler2019}. The black line shows the best log-linear fit, with the grey region representing the 90\% confidence interval calculated through bootstrapping and perturbation of the data. We find a significant positive correlation, as evaluated using the Spearman coefficient, in 86\% of trials. The best-fit slope is $0.97\pm0.36$, indicating a close-to-linear correlation between efficiency and mass.}
    \label{fig:M-eff}
\end{figure}

One of the panels of Figure \ref{fig:corner}, which we show more clearly in Figure \ref{fig:M-eff}, suggests a correlation between the radiative efficiency, $\epsilon$, and SMBH mass, $M_\bullet$. To determine the significance of the correlation, we carry out a series of Monte Carlo tests. For 10,000 iterations, we draw a bootstrapped (random with replacement) sample of 32 objects, and then resample the values of each parameter using draws from Gaussian distributions with widths $\sigma^2 = \sigma^2_{\rm stat} + \sigma^2_{\rm sys}$ (this is the "bootstrapping-plus-perturbation" method of \citealt{Curran2014}). Each $\sigma_{\rm stat}$ is our statistical error on $\log M_\bullet$ or $\log \epsilon$, while $\sigma_{\rm sys}=0.2$ or 0.68\,dex are their respective systematic errors \citep[estimated by][]{Mockler2019}.  At each iteration, we calculate the Spearman's rank correlation coefficient ($\rho$) and its significance, along with the slope of the best-fit straight line from least-squares regression. We find $\rho=0.41\pm0.16$, with $p_{\rm null}<0.05$ in 62\% of draws. The slope of the relation corresponds to a power-law $\epsilon \propto M_\bullet^{0.97\pm0.36}$, i.e.~consistent within the errors with a linear relation.

Having established the significance of this correlation, we next check for a degeneracy between these parameters by inspecting the 2D posteriors of every TDE in the sample. In checking these posteriors, we observe that $\epsilon$ is usually degenerate with the mass of the disrupted star, as found by \citet{Mockler2021}, but generally weakly or not at all with $M_\bullet$ (see example in Figure \ref{fig:excorn}). We therefore disfavour degeneracy as a likely cause of the observed correlation, but cannot fully rule it out.

If the correlation is real, it may provide an interesting constraint on the SMBHs that host TDEs. Radiative efficiency is typically linked to the aligned spin of the SMBH, with greater efficiency for larger spin magnitudes and prograde orbits \citep[e.g.][]{Jonker2020}. Higher efficiencies in systems with more massive SMBHs could suggest that the more massive objects are rotating more rapidly, and/or hosting more prograde TDEs. This picture is complicated at present by our uncertainty in the TDE emission mechanism and its dependence on SMBH mass. For example, a scenario in which TDEs in the low-mass regime are powered mainly by stream self-intersection shocks, while those in the high-mass regime are powered mainly by accretion, may also produce such a correlation.

We also note an apparent correlation between the two photospheric parameters, $R_{\rm ph,0}$ and $l_{\rm ph}$, but a physical meaning to this correlation is difficult to distinguish from parameter degeneracy, since these variables together control the size of the photosphere. This degeneracy is visible in the example 2D posteriors shown for a representative TDE in Appendix \ref{sec:caveats}.

\subsection{Implications for TDE diversity}

The systematic difference in the scaled impact parameter, $b$, between H and He/H+He TDEs is an important clue to their spectroscopic diversity. Most of the TDE-H+He events cluster around $b=0.8-1$ (and TDE-He slightly higher), corresponding to near-complete disruption of the star, while the TDE-H class prefers lower values corresponding to partial disruptions. Comparing the impact parameters in the TDE-H population with the partial TDE simulations of \citet{Ryu2020III}, we expect $\sim25$\% of a typical $\sim0.5$\,\M\ star would survive as a bound remnant following the encounter.

This suggests one potentially simple factor that could affect TDE spectroscopic diversity. Full disruptions have access to mass from inside the stellar core, providing a greater source of nuclear-processed material rich in helium (and potentially also N and O). In contrast, the partial disruptions strip only the hydrogen envelopes of the stars. However, without knowing how far these stars have progressed through their main sequence evolution, and hence the fraction of mass in the helium core, this explanation remains tentative, and is complicated by the three TDE-H+He in our sample that do prefer partial disruptions. Moreover, if TDE progenitors are born in the most recent star-formation episode in their host galaxies, typical ages are likely $\lesssim 1$\,Gyr \citep{French2020a} -- relatively young for these low-mass stars, so possibly disfavouring helium abundance in the debris as an important factor.

{\renewcommand{\arraystretch}{1.5}
\begin{table*}
    \centering
        \caption{Differences between the TDE-H and TDE-H+He (and TDE-He) populations.}
    \begin{tabular}{c|c|c|c|c}
        Property & TDE-H & TDE-He/H+He & Strength of evidence & Ref \\
         \hline
        Spectrum & Hydrogen & (H\,I), He\,II, Bowen & Observational definition & \citet{vanVelzen2021} \\
        Radius   & $>10^{15}$\,cm & few\,$\times10^{14}$\,cm & $p<0.001$ (KS) & \citet{vanVelzen2021} \\
        Line widths & $\gtrsim10^4$\,\kms & $\lesssim10^4$\,\kms &  $p=0.04$ (Binomial) & \citet{Charalampopoulos2021} \\
        Impact parameter & $\sim 0.9$ & $\gtrsim 1.1$ & $p=0.001$ (KS) & This work \\
        Black hole mass & $\sim 10^{6.4}$\,\M & $\sim 10^{6.8}$\,\M & Marginal & This work \\
        X-ray fraction & 3/12 events & 7/15 events & $p=0.09$ (Binomial) & This work \\
        Mechanism & Collision-induced outflow? & Prompt accretion? & Suggested interpretation & This work \\
         \hline
    \end{tabular}
    \label{tab:diffs}
\end{table*}
}

\begin{figure}
    \centering
    \includegraphics[width=\columnwidth]{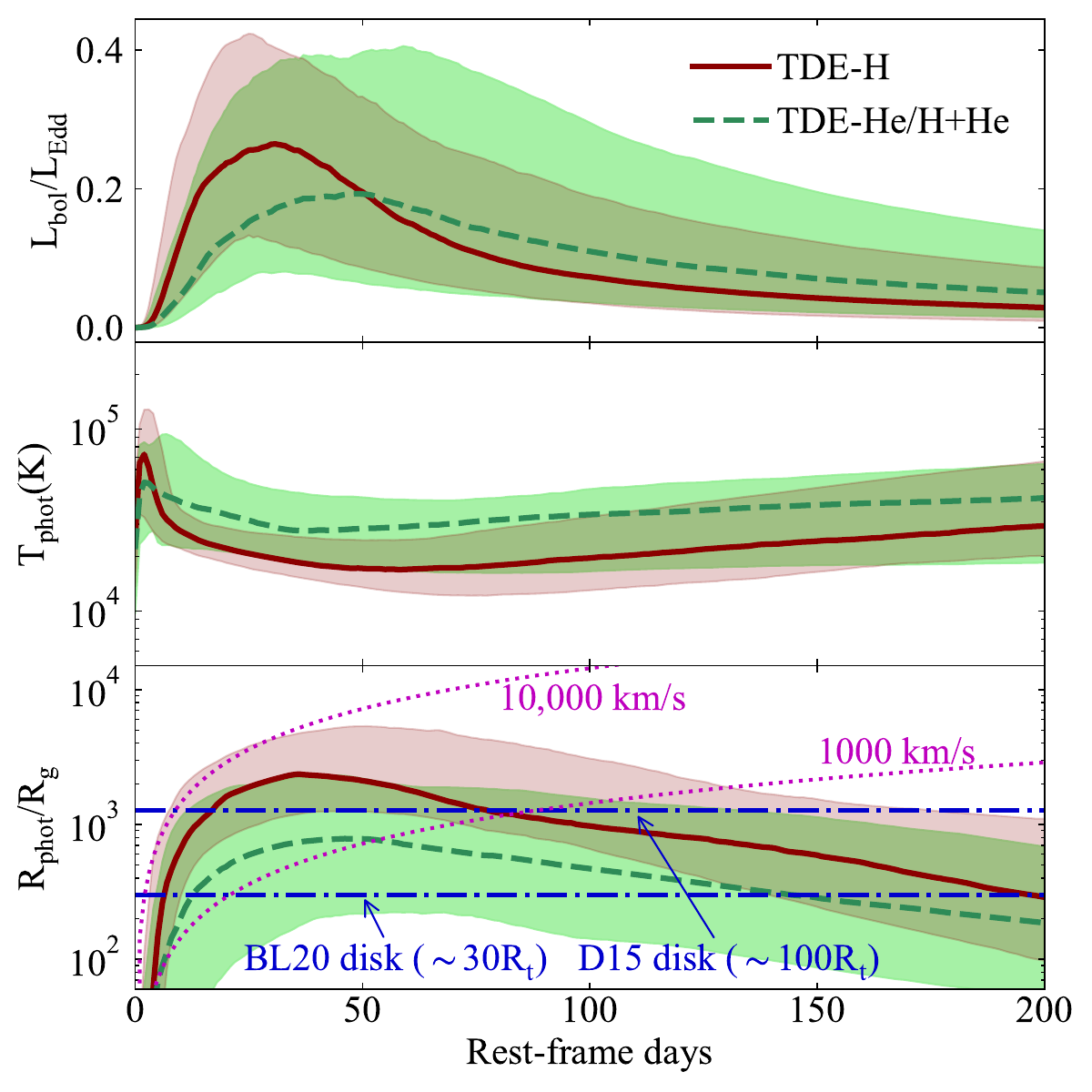}
    \caption{{Evolution of the luminosity (top), temperature (middle) and radius (bottom) for TDE-H and combined TDE-He/H+He populations. Shaded regions encompass the 16th-84th percentiles of the distributions and lines indicate the medians, calculated using 50 draws from the posterior of each TDE. Luminosities have been normalised to the Eddington luminosity and radii normalised to the gravitational radius ($G M_\bullet/c^2$) of each TDE before computing percentiles. The dotted lines in the bottom panel show the approximate size of representative outflows while the dot-dashed lines show TDE accretion disk models between $\sim{\rm few}\times 10 R_t$ \citep{Bonnerot2020} and $\sim 100 R_t$ \citep{Dai2015} for the median SMBH mass in the sample. TDE-He/H+He overlap both disks and outflows, whereas TDE-H are more consistent with fast outflows of up to $\sim 10,000$\,\kms.}}
    \label{fig:radii}
\end{figure}

The other significant difference is in the photospheric radii. Substituting {the derived set of $R_{\rm ph,0}$} into equation \ref{eq:phot}, we find that TDE-H have typical photospheric radii $\sim 10^{15}$\,cm while TDE-H+He have photospheric radii $\sim 10^{14}$\,cm. {The derived radii are in broad agreement with other photometric sample studies by \citet{vanVelzen2021a}, \citet{Hinkle2020} and \citet{Hammerstein2022}. We plot the time-dependent photospheric temperatures and radii of the TDE-H and TDE-He/H+He samples in Figure \ref{fig:radii}.} Our findings confirm that the size of the photospheric radius is an important determinant of spectroscopic type, as initially pointed out by \citet{vanVelzen2021a}. The reasons for these differences have not been readily apparent, but our modelling results provide important clues that we will now discuss. We summarise the known differences between classes, and new ones identified here, for reference in Table \ref{tab:diffs}.

In fact, the differences in photospheric radii may be linked to the differences we find in impact parameter and SMBH mass. TDE-H have average impact parameters smaller by $\sim 25\%$ and SMBH masses lower by a factor of a few. The stream self-intersection radius scales as $r_c \propto \beta^{-3} M_\bullet^{-1}$ \citep{Kochanek1994,Jiang2016}, which is $\sim 5$ times larger on average for TDE-H compared to TDE-H+He. In addition to a larger characteristic self-intersection radius, shallower encounters also lead to stream collisions at wider angles due to weaker precession (by $\sim 0.5$\,rad for typical parameters here), which result in a larger fraction of orbital energy converted to outflows \citep{Jiang2016}. These effects combined could naturally produce an envelope an order of magnitude larger in TDE-H events. Moreover, faster outflows could explain the broader spectral lines widths in the TDE-H population, identified by \citet{Charalampopoulos2021}.

Deeper encounters \citep{Jiang2016} and more massive SMBHs \citep{Guillochon2015}, in contrast, are expected to hold onto more material, and lead to self-intersection closer to the SMBH \citep{Dai2015}, forming accretion disks more rapidly. This would help to account for the spectroscopic evidence (Bowen lines) for rapid-onset  accretion in the TDE-H+He class. An appealing scenario, therefore, is one in which the emission from most TDE-H events is powered by stream collisions, reprocessed in extended collision-induced outflows \citep[see also][]{Lu2020}, while TDE-H+He are powered by accretion. This accretion luminosity must also be reprocessed to generate the observed temperatures, but this would occur in a more compact atmosphere or accretion-driven wind \citep{Metzger2016}. The higher densities in TDE-H+He events would also increase the efficiency of the Bowen fluorescence mechanism, as pointed out by \citet{vanVelzen2021}.

Circumstantial evidence to support this scenario also comes from the skew towards high radiative efficiencies observed in the TDE-H+He population, but not in the TDE-H population (Figure \ref{fig:corner}). While the difference in $\epsilon$ is not statistically significant between these populations, it is interesting when considered in light of our observed correlation between $\epsilon$ and $M_\bullet$. Accretion is typically expected to be more efficient than stream collisions at converting energy to radiation. Prompt disk formation around more massive SMBHs could explain this correlation, and in that case would also account for the preference exhibited by TDE-H+He towards massive SMBHs.

Applying the same logic, TDE-He events have the largest impact parameters and SMBH masses, and hence their streams self-intersect at the smallest radii. As suggested by \citet{Nicholl2019} and \citet{Charalampopoulos2021}, if these systems are the most compact, the lack of hydrogen lines can be explained by the \citet{Roth2016} model, in which Balmer lines are self-absorbed if the emitting region is too small.

{The \citet{Roth2016} model for spectral line formation requires an optically thick, electron scattering photosphere. Moreover, our \textsc{mosfit} model also assumes thermalisation at the scale of the photosphere \citep{Mockler2019}. A simple test shows that this assumption is reasonable even in the case of the shallow encounter in a TDE-H. For disrupted masses $\gtrsim 0.1$\,\M, and our measured photospheric radii, we find that the debris from a TDE-H (TDE-H+He) has an average density $\approx {\rm few}\times 10^{-14}\,(10^{-11})$\,g\,cm$^{-3}$, giving an optical depth $\tau \sim 10$ (1000), assuming solar composition with opacity $\kappa=0.34$\,cm$^2$\,g$^{-1}$. This assumes spherical symmetry, which is an important caveat; a more complex debris geometry could increase the optical depth along some sightlines and decrease it along others.}

{
Another important test for powering mechanisms comes from the X-rays. While accretion is expected to produce substantial X-ray emission from the inner disk (at least for favourable viewing angles), it is less obvious that emission due to stream intersection should be luminous at X-ray wavelengths. Our sample includes 10 TDEs with reported X-ray detections, of which three are TDE-H and seven are TDE-He or H+He. These events are highlighted in Table \ref{tab:tdes}. Overall 10/27 spectroscopically classified TDEs are X-ray bright. Under the null hypothesis that X-ray TDEs show no preference for spectroscopic type, a binomial test tells us that probability of finding X-rays in 3/12 TDE-H and in 7/15 TDE-He/H+He is $p=0.09$. While at face value X-rays have been detected roughly twice as often for TDEs with He~II, potentially supporting the idea that these events are more likely to have experienced rapid disk formation, larger samples will clearly be needed to establish whether this is significant. Noteably, the presence of X-rays in some TDE-H indicates either that at least some of these events also commence accretion rapidly or that stream-collisions can also produce X-ray emission.
}

While we caution that the Other group in our study may not represent a coherent class due to lack of spectroscopic data for some events, these objects clearly stand out in terms of their photospheres, which are more compact and less sensitive to luminosity than the other classes. Their smaller photospheres may be responsible for the lack of spectral lines in some featureless events, due to ionization \citep{Guillochon2014} or self-absorption \citep{Roth2016} effects. However, in terms of their other parameters, we find that these events are broadly similar to the TDE-H+He class.

We do not find any events requiring very deep ($b \gg 1$) encounters. We speculate that this may result from selection effects. For example \citet{Gafton2019} show that for close encounters, the peak value of the mass fallback rate, and/or the time to reach this peak, decreases. Such TDEs would be detectable over a smaller volume and would be less likely to be classified by current searches, which often select for slow-rising transients. Future searches may wish to target fainter and faster events, though we note that the faintest and fastest TDE in our sample, iPTF16fnl, does not require an especially large impact factor.

{\citet{vanVelzen2021} offered an alternative explanation for the difference in photospheric radii between the TDE subclasses. Assuming that all TDEs have $\beta\approx1$ and form accretion disks, the disk radius scales with $M_*$ (for a given $M_\bullet$) \citep{Dai2015}. If the radius of the photosphere is proportional to the disk radius, then the mass of the disrupted star would be the key parameter that differs between classes. Evidence for this picture comes from the relatively lower volumetric rate of the TDE-H population inferred by \citet{vanVelzen2021} -- a larger photosphere and a lower rate would both be expected to arise for more massive stars. However, our modelling suggests no differences in the mass of the disrupted stars. Moreover, the inferred sizes of photospheres for the TDE-H class exceed that available from disk models (Figure \ref{fig:radii}. \citet{Krolik2020} find that partial disruptions should occur roughly 1/3 as often as `common' full disruptions, roughly consistent with the difference in rates between TDE-H and TDE-H+He. However, they found that prompt circularisation occurred only for deeper encounters than our fits here. It may be that both stellar mass and impact parameter play important roles in determining the spectroscopic character of a given TDE.}

\section{Conclusions}
\label{sec:conc}


Motivated by the emergence of spectroscopic sub-classes within the growing statistical sample of TDEs, we have modelled the light curves of 32 well-observed TDEs with \textsc{mosfit} to look for differences in the inferred physical parameters and whether these correlate with spectroscopic type. We provide a catalog of parameters from our modelling in Table \ref{tab:tdes}.

We find SMBH masses in the range $\approx 10^{5.8}-10^{7.2}$\,\M, and a relatively shallow distribution of disrupted star masses, in the range $\approx 0.1-1$\,\M, compared to a fiducial IMF. The population of disrupted stars does not appear to differ between the TDE spectroscopic types, and none of the model fits require the disruptions of stars below the hydrogen burning limit. While the mass distribution of observed TDEs may be influenced by selection effects, it also likely reflects the fact that for lower mass stars there is a narrower range of SMBH masses that can disrupt these stars outside of the event horizon.

We identify a strong correlation in our sample between radiative efficiency and SMBH mass. This could suggest that more massive SMBHs host more prograde TDEs (or rotate faster), or support a difference in powering mechanism at the high (accretion) and low (stream collision) ends of the $M_\bullet$ distribution.

We find evidence for several systematic differences between TDE-H and TDE-H+He events:
\begin{itemize}
    \item TDE-H have statistically smaller values of the scaled impact parameter $b$, corresponding to shallower encounters and less complete disruptions (with TDE-He the most complete)
    \item TDE-H also have significantly larger photospheric radii, consistent with results from \citet{vanVelzen2021}, \citet{Charalampopoulos2021}, and \citet{Hinkle2020}
    \item TDE-H+He show moderate evidence for more massive SMBHs on average
    \item TDE-H+He skew toward higher radiative efficiencies, with the median larger by 0.4\,dex (this effect is not statistically significant, but interesting given the correlation of $\epsilon$ with $M_\bullet$).
\end{itemize}

The systematic differences in photosphere parameters are consistent with existing theories of spectral line formation. The compact photospheres in TDE-H+He encourage the high densities required for the Bowen fluorescence mechanism observed in this class. The requirement for extended photospheres in TDE-H events is consistent with the \citet{Roth2016} models in which Balmer lines are self-absorbed if the photosphere is too small. However, our novel finding is the identification of differences in the impact parameter (and possibly SMBH mass) as a potential cause of the differences in photospheric radii.

One possibility that could account for several of the above differences is that a typical TDE-H is powered by the collisionally-induced outflow launched by the intersection of debris streams. The smaller impact parameter and lower SMBH mass results in the streams colliding at larger radii and wider angles, leading to larger characteristic photospheres and lower radiative efficiency. The corollary is that deeper and fuller disruptions around more massive SMBHs pack their debris densely and begin accreting promptly \citep[e.g.][]{Krolik2020}, leading to the Bowen fluorescence lines in a TDE-H+He spectrum as well as higher radiative efficiency. 

Of course, in reality this picture will not be so clear cut, as delayed onset accretion (and likely numerous other factors) will blur the boundaries between these classes. While indicative for now, our statistical results and the estimated physical parameters we provide for these events will be important in understanding TDE diversity. With larger samples on the horizon, in particular from LSST \citep{Bricman2020}, this kind of analysis will help to determine the (potentially different) energy generation mechanisms in each of the TDE classes, and as a function of SMBH mass.

\section*{Acknowledgements}

We thank an anonymous referee for their helpful comments to improve the manuscript. We are grateful for feedback from Sjoert van Velzen and Julian Krolik, and further thank Sjoert van Velzen for providing ZTF light curve data. MN acknowledges support from the European Research Council (ERC) under the European Union’s Horizon 2020 research and innovation programme (grant agreement No.~948381) and a Fellowship from the Alan Turing Institute.

\section*{Data Availability}

This work is based on publicly available data from the Open TDE Catalog \citep{Guillochon2017} and from \citet{vanVelzen2021}.



\bibliographystyle{mnras}
\bibliography{refs} 




\appendix

\section{Priors, limitations and differences with Mockler et al.~2019}
\label{sec:caveats}

\begin{table}
  \caption{Priors for the \textsc{mosfit} TDE model. Priors are flat within the stated ranges, except for $M_*$, which uses a Kroupa initial mass function.}
  \centering
  \begin{tabular}{ccc}
  \hline
  Parameter & Prior & Units\\
  \hline
$ \log{(M_\bullet )}$ & $[5, 8]$  & M$_\odot$ \\
$ M_*$ & $[0.01, 100]$ & M$_\odot$ \\
$ b$ & $[0, 2]$ &    \\
$ \log(\epsilon) $ & $[-4, -0.4] $  &   \\
$ \log{(R_{\rm ph,0} )} $ & $[-4, 4] $  &   \\
$ l_{\rm ph}$ & $[0, 4]$ &    \\
$ \log{(T_v )} $ & $[-3, 3] $  & days  \\
$ t_0 $ & $[-500, 0]$ &  days  \\
$ \log{(n_{\rm H,host})}$ & $[19, 23]$  & cm$^{-2}$ \\
$ \log{\sigma} $ & $[-4, 2] $  &   \\
  \hline
\end{tabular}
  \label{tab:priors}
\end{table}

\begin{figure*}
    \centering
    \includegraphics[width=\textwidth]{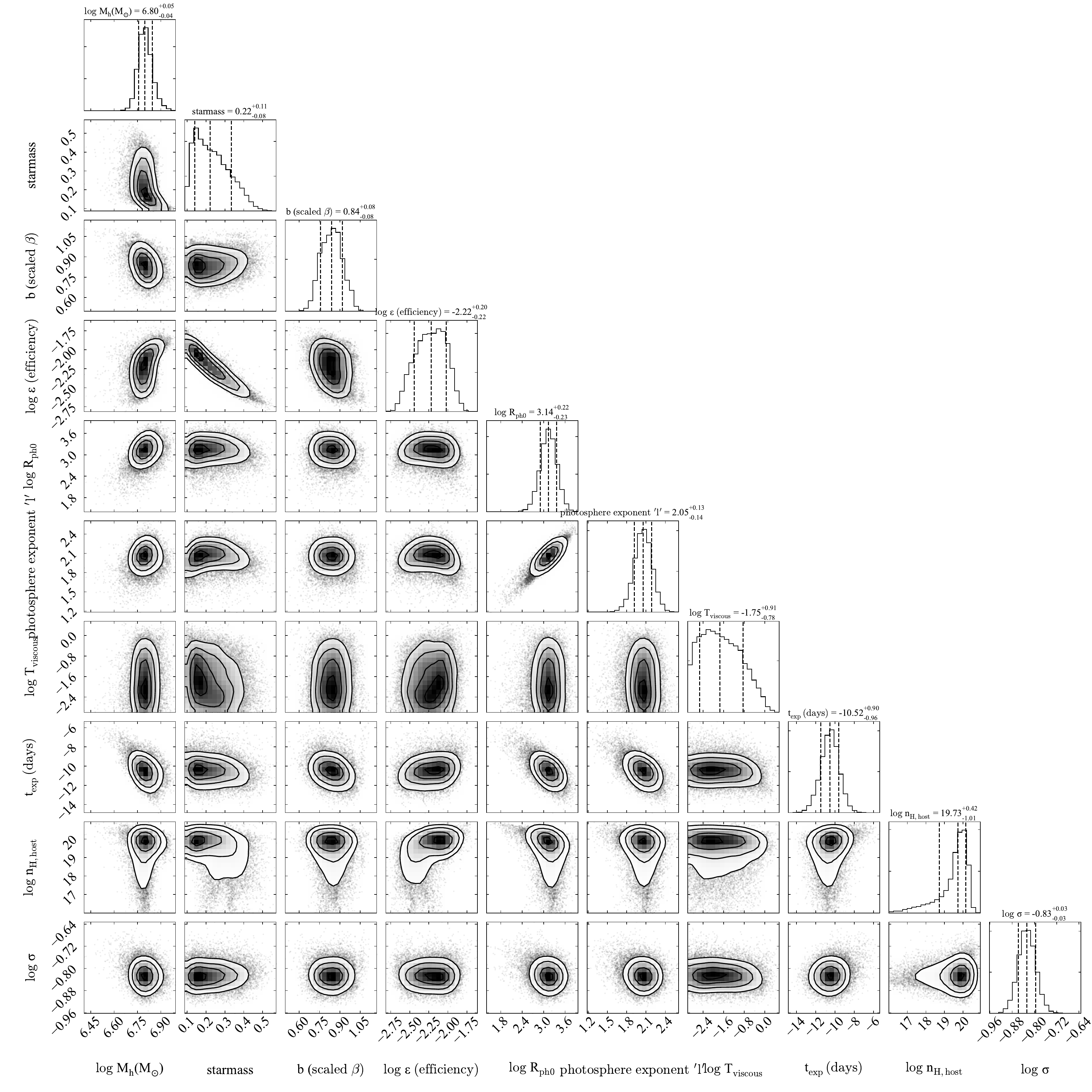}
    \caption{Example of a typical corner plot, showing 2D posteriors for AT2018zr.}
    \label{fig:excorn}
\end{figure*}

{
The priors used in our fits are listed in Table \ref{tab:priors}. These are mostly the same as in \citet{Mockler2019}, however we use a broader prior on the radiative efficiency, $\epsilon$. \citet{Mockler2021} find that there is a significant degeneracy between $\epsilon$ and the mass of the disrupted star, and recommend a broader prior on $\epsilon$ to avoid biasing estimates of $M_{*}$. Two other significant differences exist between our method and that of \citet{Mockler2019}: our use of the \textsc{dynesty} sampler, rather than \textsc{emcee}, and the re-calibration of the UVOT photometry. 
}

{
The change in the UVOT data affects five TDEs studied by \citet{Mockler2019}: ASASSN-14ae, ASASSN-14li, ASASSN-15oi, iPTF16axa and iPTF16fnl. The recalibrated data give a higher UV luminosity and therefore a higher photospheric temperature. This appears to result in smaller photospheric radii in our fits to iPTF16fnl and iPTF16axa, though it does not seem to make a substantial difference for the ASASSN events. There are no obvious systematic changes in other parameters.
}

{
The changes in efficiency prior and sampler makes a bigger difference. The largest discrepancies between our results and those of \citet{Mockler2019} are in the efficiency and stellar mass parameters. Our results often find solutions at lower efficiency than was allowed in the earlier study, e.g. for PTF09ge, iPTF16axa. This has knock-on effects to the stellar mass due to the known degeneracy. For the GALEX TDEs, D1-9 and D3-13, we find more modest $M_*\sim0.3$\,\M, compared to somewhat large masses $\sim7$\,\M\ in \citet{Mockler2019}.
}

{
To test the effects of the sampler, we re-run fits to all the TDEs in our sample using \textsc{emcee}, and compare the results to those derived with \textsc{dynesty}. While overall most results are consistent, we find disagreements in cases where posteriors are multimodal. For example, ASASSN-14ae shows a bimodal posterior in $M_*$, with a narrow peak at $\lesssim 1$\M\ and a broad peak centered at $\sim 2$\M. While both peaks are visible in the posterior from \textsc{dynesty}, the narrow peak (where most of the probability mass resides) is not well explored by \textsc{emcee}. In general, nested sampling seems less susceptible to local minima and more reliably samples the full parameter space. We also note that many of the \textsc{emcee} runs return an unrealistically high stellar mass (the median for the H+He sample with \textsc{emcee} is 6.2\,\M). Other parameters, in particularly the black hole mass, are largely insensitive to the choice of sampler.
}

{
We therefore elect to focus our analysis on the \textsc{dynesty} results, but note that the derived star masses and efficiencies in particular should be treated with some caution due to their sensitivity to choices of prior and sampler.
}

Figure \ref{fig:excorn} shows the typical posteriors we obtain from \textsc{mosfit}, for a representative TDE (AT2018zr). Degeneracies between $R_{\rm ph,0}$ and $l_{\rm ph}$, and between $\epsilon$ and $M_*$ \citep{Mockler2021}, are apparent for most events. Weaker degeneracies are sometimes present between $M_\bullet$ and $b$, and also between $M_\bullet$ and the photospheric parameters. In some events, a mild degeneracy is seen between $\epsilon$ and $M_\bullet$ as seen in Figure \ref{fig:excorn}, though not always in the same direction as the correlation in Figure \ref{fig:M-eff}.


\bsp	
\label{lastpage}
\end{document}

%% file: affils.tex
\newcommand{\bham}{School of Physics and Astronomy, University of Birmingham, Birmingham B15 2TT, UK}
\newcommand{\igw}{Institute for Gravitational Wave Astronomy, University of Birmingham, Birmingham B15 2TT, UK}
\newcommand{\edinburgh}{Institute for Astronomy, University of Edinburgh, Royal Observatory, Blackford Hill, EH9 3HJ, UK}

\newcommand{\ucsc}{Department of Astronomy and Astrophysics, University of California, Santa Cruz, CA 95064, USA}
\newcommand{\dark}{DARK, Niels Bohr Institute, University of Copenhagen, Blegdamsvej 17, DK-2100 Copenhagen, Denmark}